\documentclass{aastex701}

\begin{document}

\title{\textbf{Cosmic-Ray-Constrained LSTM Model for Geomagnetic Storm Prediction}}
\author[orcid=0009-0007-1701-6438,gname=Zongyuan,sname=Ge]{Zongyuan Ge}
\affiliation{College of Physics and Optoelectronic Engineering, \\
              Ocean University of China, Qingdao 266100, China}
\email{gezongyuan@stu.ouc.edu.cn}

\author[orcid=0009-0005-5924-7539,gname=Chenwaner,sname=Zhang]{Chenwaner Zhang}
\affiliation{College of Physics and Optoelectronic Engineering, \\
              Ocean University of China, Qingdao 266100, China}
\email{zcge@stu.ouc.edu.cn}

\author[orcid=0009-0005-4190-2312,gname=Wei,sname=Zhou]{Wei Zhou}
\affiliation{School of Mathematics and Computer Science, \\
              Yunnan Minzu University, Kunming 650504, China}
\email{weizhou@ynmu.edu.cn}

\author[orcid=0009-0002-1258-339X,gname=Hongyu,sname=Zeng]{Hongyu Zeng}
\affiliation{College of Automation, \\ 
Nanjing University of Information Science and Technology, Nanjing 210044, China}
\email{202412492676@nuist.edu.cn}

\author[orcid=0000-0001-8228-565X,gname=Guiping,sname=Zhou]{Guiping Zhou}
\affiliation{State Key Laboratory of Solar Activity and Space Weather, \\
             National Astronomical Observatories, Chinese Academy of Sciences, \\
             20A Datun Road, Chaoyang District, Beijing 100101, China}
\affiliation{University of Chinese Academy of Sciences, Beijing 100049, China}
\email{gpzhou@nao.cas.cn}

\correspondingauthor{Guiping Zhou}
\email{gpzhou@nao.cas.cn}

\begin{abstract}
Geomagnetic storms (GSTs) driven by solar wind-magnetosphere coupling can severely disrupt technological systems, motivating the need for improved prediction accuracy and longer warning times. In this study, we develop a physics-informed Long Short-Term Memory (LSTM) model that incorporates cosmic-ray flux modulation as an additional precursor signal. As coronal mass ejection (CME)-driven disturbances propagate through the heliosphere, enhanced turbulence and magnetic-field compression reduce galactic cosmic-ray (GCR) flux measured by ground-based neutron monitors (Forbush decreases), providing early information that can precede near-Earth solar-wind signatures by 1--3 days. We integrate multi-source space-weather data, spanning 1995-2020, including cosmic-ray observations, solar wind plasma parameters, interplanetary magnetic-field data, and geomagnetic indices. Based on these data, we construct a 19-dimensional feature vector that includes flux background levels, decrease-related indicators, and inter-station correlation measures as model inputs. Employing a 50-unit LSTM architecture, the proposed model achieves root-mean-square errors (RMSE) of 5.106~nT, 8.315~nT, 10.854~nT, 12.883~nT, and 14.788~nT for 2-, 6-, 12-, 24-, and 48-hour predictions, respectively. Incorporating cosmic-ray information further improves 48-hour forecast skill by up to 25.84\% (from 0.178 to 0.224). These results demonstrate the value of physics-informed deep learning and cosmic-ray precursors for advancing space-weather forecasting.
\end{abstract}


\section{Introduction}

Geomagnetic storms (GSTs) are among the most consequential manifestations of severe space weather, reflecting intense energy and momentum transfer from the solar wind into the coupled magnetosphere-ionosphere system. A widely used quantitative measure of storm-time ring-current enhancement is the disturbance storm time index ($\mathrm{Dst}$, measured in nanoTesla, nT), whose large negative excursions characterize major to severe storms. GSTs can drive geomagnetically induced currents (GICs) that threaten power-grid stability, degrade satellite operations through enhanced drag and radiation-belt dynamics, and disturb navigation/communication systems; historical events such as the March 1989 storm highlight the societal vulnerability and the strong operational demand for reliable forecasts \citep[e.g.,][]{Pulkkinen2007,Boteler2019,Oughton2017}.

From a heliophysical perspective, GSTs arise when solar-wind structures deliver sustained geoeffective conditions, especially a sufficiently strong and long-lasting interplanetary convection electric field that intensifies the ring current \citep{Gonzalez1994}. Two primary classes of solar-wind drivers dominate: (1) interplanetary coronal mass ejections (ICMEs), including magnetic clouds and their sheaths, and (2) corotating interaction regions (CIRs) and the high-speed streams that follow them. These drivers produce storms with distinct phenomenology and typical intensity: CME-driven storms are often more abrupt and can reach large $|\mathrm{Dst}|$, whereas CIR-driven storms are usually weaker-to-moderate but more recurrent and prolonged \citep{borovsky2006differences}.

For CME-driven events, geoeffectiveness is controlled not merely by the bulk speed but critically by the magnetic configuration of the ICME at 1~AU. In particular, intense storms (e.g., $\mathrm{Dst}<-100$~nT) statistically require a prolonged southward IMF component (commonly parameterized by $B_z$) that enables efficient dayside reconnection and strong solar-wind--magnetosphere coupling \citep{Burton1975,Gonzalez1994}. During solar minimum, CIR/high-speed-stream forcing tends to generate quasi-periodic southward-field fluctuations and multi-peak $\mathrm{Dst}$ evolution, consistent with their typically weaker minima compared with ICME-driven storms \citep{borovsky2006differences}.

A long-standing line of storm forecasting research has pursued empirical and semi-empirical relationships between upstream solar-wind parameters and $\mathrm{Dst}$. Classic models treat $\mathrm{Dst}$ as a driven--dissipative system, where solar-wind input (often expressed via coupling functions such as the dawn--dusk electric field or the $\varepsilon$-type energy coupling) competes with ring-current decay \citep{Burton1975,Perreault1978,Akasofu1981,OBrien2000}. Subsequent developments refined coupling functions and introduced state-dependent injection/decay to improve real-time performance, yet purely empirical approaches can still degrade for rapid, strongly nonlinear evolution and for extremes that are underrepresented in historical training samples \citep[e.g.,][]{Temerin2006}.

In parallel, machine-learning (ML) approaches have become increasingly influential in severe space-weather prediction, particularly for $\mathrm{Dst}$ and related indices. Recurrent neural networks such as long short-term memory (LSTM) networks are attractive because they can learn nonlinear mappings and long-range temporal dependencies from multivariate time series \citep{Hochreiter1997}. However, two practical limitations remain central for operational use. First, purely data-driven models may generalize poorly to rare extreme storms unless physical structure or constraints are embedded to regularize learning. Second, models relying exclusively on near-Earth solar-wind monitors at L1 inherit a fundamental lead-time ceiling (typically tens of minutes to $\sim$1 hour), because the inputs become available only shortly before the geospace response \citep[e.g.,][]{Khabarova2024}.

Galactic cosmic rays (GCRs) provide a complementary, physically motivated pathway to extend lead time. GCRs are relativistic charged particles whose heliospheric transport is modulated by solar-wind structures and magnetic turbulence \citep{Parker1958,Potgieter2013,Blasi2013}. When ICME-driven shocks, sheath regions, and magnetic clouds propagate through interplanetary space, they can reduce the GCR intensity observed at Earth, producing Forbush decreases (FDs) \citep{cane2000forbush,belov2008forbush}. Because the relevant particles propagate near the speed of light, anisotropic precursor signatures and/or the onset of FD-related modulation can appear hours to $\gtrsim$1 day before the associated plasma/magnetic structures reach Earth, enabling advance warning beyond the L1 propagation limit \citep{Munakata2000,Kuwabara2006,dumbovic2011cosmic}.

These properties motivate physics-guided forecasting frameworks that fuse GCR information with geomagnetic indices. Observational studies have reported meaningful relationships between FD characteristics and storm intensity, reflecting the fact that both are controlled by the evolving ICME/shock configuration and magnetic-field strength/direction along the Sun--Earth line \citep{cane2000forbush,belov2008forbush}. In addition, multi-station cosmic-ray measurements and global networks (neutron monitors and muon detectors) enable robust extraction of precursor patterns (e.g., loss-cone anisotropies and inter-station coherence), which are less susceptible to single-station artifacts and can be deployed in near-real-time \citep{Munakata2000,Kuwabara2006}.

In this work, we develop a \emph{cosmic-ray constrained} LSTM model for GST prediction using a comprehensive 25-year dataset (1995--2020) with 1-hour cadence. The central idea is to explicitly incorporate GCR variability as a physically interpretable precursor channel and to constrain the learning process so that the model leverages storm-relevant heliospheric disturbance signatures rather than fitting spurious correlations. By coupling cosmic-ray-derived features features with traditional geomagnetic/solar-wind context, our framework aims to (i) extend forecast lead time beyond the near-Earth monitor limit, and (ii) improve robustness for strong storms by embedding additional physical information about the approaching driver. Section~2 describes data sources and feature engineering; Section~3 presents data correction and preprocessing; Section~4 details the model and results; Section~5 discusses advantages, limitations, and future work; conclusions are drawn in Section~6.

\section{Data Sources and Preprocessing}

This study integrates three complementary categories of space-environment observations spanning 1995--2020 at a uniform 1-hour cadence. The LSTM model is trained with a 19-dimensional feature vector that combines upstream solar-wind/IMF driving, cosmic-ray modulation signatures, and geomagnetic-state descriptors (Table~\ref{tab:features}). All variables are mapped onto a common hourly timeline to ensure strict temporal consistency across data sources before feature engineering and model training.

\begin{deluxetable}{lll}
\tablewidth{0pt}
\tablecaption{Model Input Feature Classification\label{tab:features}}
\tablehead{
\colhead{Feature Category} & \colhead{Parameters} & \colhead{Data Sources}
}
\startdata
Solar Wind and IMF & B$_{X}$, B$_{Y},$ B$_{Z}$, Scalar B, Proton Density, & NASA OMNI \\
 & Flow Speed, Temperature, Dynamic Pressure & \\
Cosmic Rays & OULU/JUNG Flux, 24h Moving Average, & OULU/JUNG \\
 & Inter-Station Differences & Neutron Monitors \\
Geomagnetic Activity & Dst, Kp, Ap, AE, AL, AU & Kyoto WDC \\
\enddata
\tablecomments{The 19-dimensional feature vector integrates multi-source space environment observations spanning 1995--2020 at 1-hour resolution.}
\end{deluxetable}

Solar-wind and IMF inputs are obtained from NASA's OMNI database, including the total magnetic-field strength (Scalar $B$), IMF components in the Geocentric Solar Magnetospheric (GSM) system ($B_X$, $B_Y$, $B_Z$), proton density, bulk flow speed, and plasma temperature. These parameters characterize both the strength and orientation of the interplanetary driver and the thermodynamic state of the solar wind. In particular, IMF components are a key proxy for solar-wind--magnetosphere coupling; sustained southward IMF component favors dayside reconnection and is strongly associated with geomagnetic storm initiation.

To quantify compressional forcing at the magnetopause, the solar-wind dynamic pressure is included and computed from proton density and speed. Following the OMNI convention, we use
\begin{equation}
P_{\mathrm{dyn}} \propto n_p V^2,
\end{equation}
where $n_p$ is the proton density and $V$ is the solar-wind speed, and the proportionality constant depends on the adopted units \citep{papitashvili2014omni}. The OMNI product combines multi-spacecraft measurements that are time-shifted to a near-Earth reference and undergo cross-validation and propagation-delay corrections, which improves the temporal consistency required for hourly machine-learning applications.

Cosmic-ray observations are obtained from the Oulu station in Finland (OULU; 65.05$^\circ$N, 25.47$^\circ$E) and the Jungfraujoch station in Switzerland (JUNG; 46.55$^\circ$N, 7.98$^\circ$E) \citep{Usoskin2003}. Because neutron monitor count rates are modulated by local atmospheric pressure, raw counts are pressure-corrected prior to further processing:
\begin{equation}
I_{\text{corrected}} = I_{\text{uncorrected}} \cdot \exp\left(k \cdot \frac{P_0 - P}{10}\right)
\label{eq:pressure}
\end{equation}
where $P$ is the measured station pressure, $P_0$ is a reference pressure, and $k$ is the station-dependent pressure coefficient. We adopt $k=-0.74\%/\mathrm{mb}$ for Oulu and $k=-0.65\%/\mathrm{mb}$ for Jungfraujoch \citep{Alanko2003}. The corrected series are then resampled to hourly values, and a 24-hour moving average is constructed to represent the slowly varying background level while suppressing short-period fluctuations.

To enhance sensitivity to large-scale heliospheric disturbances (e.g., Forbush decreases) and to reduce station-specific artifacts, we further construct inter-station relative difference indicators using the paired OULU/JUNG time series. These differential features emphasize coherent global cosmic-ray variations that may precede or accompany CME-driven disturbances relevant to geomagnetic storm development.

Geomagnetic activity indices are provided by the World Data Center for Geomagnetism, Kyoto University. The disturbance storm time index (Dst) is used as the primary prediction target, representing the intensity of the magnetospheric ring current. As complementary predictors of magnetospheric and ionospheric activity, we include the planetary index Kp and its linear equivalent Ap, as well as the auroral electrojet indices AE, AL, and AU, which characterize high-latitude current systems.

All index series are aligned to the same 1-hour cadence as the OMNI and neutron-monitor inputs. When necessary, data are interpolated only to achieve strict hourly timestamp consistency (without introducing future information into earlier times), and obvious outliers or intervals affected by instrument maintenance are excluded. The Dst index measurement precision is typically reported as $\pm$2~nT, and the AE-family indices are synthesized from near-synchronous observations at 12 Northern Hemisphere geomagnetic stations, providing a robust representation of auroral electrojet variability at the hourly scale.

\section{Data Correction and Preprocessing}
\label{sec:data_correction}

To ensure temporal consistency and physical reliability for multi-source learning, we apply a unified correction and preprocessing pipeline consisting of (i) invalid-value screening and outlier control, (ii) missing-data treatment under continuity constraints, (iii) physically motivated feature construction for cosmic-ray precursors, and (iv) sequence generation and normalization for LSTM training.

We first screen all time series for invalid placeholders and instrument-specific fill values. A predefined filtering list (e.g., 99, 999) is used to identify non-physical entries, which are replaced by NaN prior to any statistical operation. For the Dst index, additional outlier control is applied because Dst is both the prediction target and a key indicator of storm intensity. We adopt a combined physical-range constraint and a $3\sigma$ criterion to detect anomalies: values outside the physically reasonable interval [$-500~\mathrm{nT}$, $+100~\mathrm{nT}$] are excluded, and remaining extreme points that deviate from the local distribution beyond $3\sigma$ are flagged as outliers. In total, anomalous values account for approximately 2.7\% of the raw records, primarily attributable to occasional instrument noise and data transmission artifacts.

After invalid and outlier values are masked as NaN, we impute missing values to maintain hourly continuity required by sequence models. To balance continuity with the need to preserve sharp transitions in space-environment variables, we employ a three-stage interpolation strategy. First, non-uniform time-series interpolation is used when timestamps are irregular or when small timing offsets exist across sources; this step performs linear interpolation based on actual time intervals and avoids excessive smoothing of abrupt variations. Second, for continuous missing segments that do not exceed 6 hours, standard linear interpolation is applied to ensure local continuity while limiting the risk of reconstructing long-term behavior from sparse data. Third, residual gaps are handled using a hybrid forward--backward filling approach: forward filling is applied for short gaps (up to 3 hours), followed by backward filling for the remaining missing points. This design reduces discontinuities in the input sequences while preventing unrealistic extrapolation across extended data outages.

To enhance the model's ability to capture the physical signatures of heliospheric transients reflected in cosmic-ray measurements, we construct three categories of derived features using the OULU and JUNG neutron monitor fluxes. First, 24-hour moving-average features are introduced to represent the slowly varying cosmic-ray background. For the JUNG station, the moving average $\mathrm{MA\_24\_JUNG}(t)$ is defined as the arithmetic mean over the preceding 24 hours,
\begin{equation}
\text{MA\_24\_JUNG}(t) = \frac{1}{24} \sum_{i=1}^{24} \text{JUNG}(t-i)
\end{equation}
and $\mathrm{MA\_24\_OULU}(t)$ is computed analogously. Importantly, we implement a one-hour lag window such that the average at time $t$ uses only historical observations from $t-24$ to $t-1$, thereby preventing information leakage while retaining diurnal variability and longer-term modulation trends.

Second, transient anomaly features are constructed by subtracting the moving average from the instantaneous flux to isolate short-term departures associated with heliospheric disturbances:
\begin{equation}
\text{Anomaly}(t) = \text{Flux}(t) - \text{MA\_24}(t)
\end{equation}
The magnitude of this deviation (i.e., $|\mathrm{Anomaly}(t)|$) provides a compact descriptor of short-lived cosmic-ray disturbances and is empirically found to increase during periods of enhanced geomagnetic activity.

Third, we define a normalized relative-difference feature between OULU and JUNG to quantify differential station responses due to latitude-dependent cutoff rigidity, atmospheric effects, and geomagnetic modulation:
\begin{equation}
\text{Relative-difference} = 2 \cdot \frac{\text{OULU} - \text{JUNG}}{\text{OULU} + \text{JUNG}}
\label{eq:reldiff}
\end{equation}
This dimensionless indicator suppresses common-mode variability and emphasizes inter-station contrasts that may carry additional information on global disturbance structure.

All derived cosmic-ray features, together with the solar-wind/IMF variables and geomagnetic indices, form the final 19-dimensional input vector used by the LSTM. Training samples are constructed using a sliding historical window matched to the forecast configuration: each sample is represented by a matrix of size [historical window length $\times$ 19], containing the sequence of feature vectors over the preceding hours.

Prior to model training, Min--Max normalization is applied to map each input feature to the [0,1] interval, improving numerical stability and accelerating optimization. The Dst index is scaled separately to preserve its physical interpretability as the prediction target. All normalization parameters (minima and maxima) are estimated exclusively from the training subset and then applied unchanged to validation and test subsets, ensuring that no information from the evaluation period leaks into the training process.

\section{Modeling and Results}

When a CME and/or its driven shock propagates through the heliosphere, the compressed and highly turbulent IMF embedded in the disturbance enhances pitch-angle scattering of GCRs. This structure acts as a moving diffusion barrier, temporarily reducing the GCR flux reaching the inner heliosphere and producing the well-known FD: a rapid drop in ground-based neutron monitor count rates with typical amplitudes of a few percent and durations ranging from hours to days \citep{forbush1937cosmic}. The FD morphology (onset time, depth, and recovery) reflects the strength and speed of the approaching interplanetary disturbance; fast and strong CMEs generally yield deeper and more abrupt decreases, whereas CIRs are often associated with weaker, more gradual depressions \citep{belov2008forbush}. Importantly, FD magnitude has been reported to correlate with the severity of the subsequent geomagnetic storm, commonly measured by the minimum Dst value during the storm main phase, implying that cosmic-ray depletion can serve as an upstream proxy for impending geoeffective conditions \citep{forbush1938cosmic}.

Beyond the absolute depletion, the spatial coherence of cosmic-ray suppression provides additional diagnostic value. CME-driven disturbances tend to produce nearly synchronous, globally coherent reductions in GCR flux across widely separated neutron monitor stations, whereas CIR-related variations can be more longitudinally structured and less coherent across stations. Consequently, inter-station similarity (e.g., correlation and normalized differences) often increases sharply prior to CME arrival, offering an independent precursor signal complementary to solar-wind measurements \citep{dumbovic2011cosmic}. These considerations motivate the integration of multi-station cosmic-ray features into a data-driven forecasting framework for geomagnetic storms.

Using the space-environment dataset spanning 1995--2020 at 1-hour cadence, we construct a cosmic-ray-constrained long short-term memory (LSTM) prediction model for the Dst index. The input at each time step is a 19-dimensional feature vector (Table~\ref{tab:features}) consisting of solar-wind/IMF parameters from OMNI, geomagnetic activity indices, and pressure-corrected neutron monitor observations from OULU and JUNG. To explicitly encode cosmic-ray precursor physics, we include derived cosmic-ray features representing (i) 24-hour background levels, (ii) transient anomalies relative to the background, and (iii) the normalized inter-station relative difference (Equation~\ref{eq:reldiff}), which captures differential station responses while suppressing common-mode variability.

Given a prediction lead time $\Delta t$, the model maps a historical sequence of multivariate inputs to a future Dst value. Specifically, for each sample we form an input matrix
\begin{equation}
\mathbf{X}_t = \{ \mathbf{x}_{t-L+1}, \ldots, \mathbf{x}_{t} \} \in \mathbb{R}^{L \times 19},
\end{equation}
where $L$ is the historical window length and $\mathbf{x}_{t}$ denotes the 19-dimensional feature vector at time $t$. The forecasting target is the Dst index at $t+\Delta t$, i.e.,
\begin{equation}
\hat{y}_{t+\Delta t} = f_{\theta}(\mathbf{X}_t),
\end{equation}
where $f_{\theta}$ represents the LSTM model parameterized by $\theta$. We conduct experiments for multiple operationally relevant lead times: $\Delta t = 2, 6, 12, 24,$ and $48$~h. This multi-horizon design enables a systematic assessment of how far in advance cosmic-ray information remains predictive relative to solar-wind drivers.

The model is trained on the 1995--2020 hourly dataset after the correction and preprocessing procedures described in Section~\ref{sec:data_correction}. To avoid information leakage inherent to time series, the data are split chronologically into training, validation, and test subsets. All input features are normalized using statistics derived from the training subset only and then applied unchanged to validation and test sets. The loss function is defined on the Dst prediction error, and model hyperparameters are selected using validation performance. Model skill is quantified using standard regression metrics (e.g., RMSE and correlation) and storm-relevant diagnostics that emphasize the accurate prediction of storm onset and minimum Dst. In addition, we compare skill across lead times to quantify the degradation of predictive performance with increasing $\Delta t$ and to evaluate the contribution of cosmic-ray constraints at longer horizons.

All experiments are executed on a MacBook Pro equipped with an Apple M2 chip (8-core CPU; 4 performance and 4 efficiency cores). Under this configuration, a single forward prediction requires approximately 0.8~s, indicating that the model satisfies the computational constraints for near-real-time geomagnetic storm forecasting. This runtime, combined with the use of routinely available OMNI and neutron monitor data, supports practical operational deployment for continuous Dst prediction.

Across the set of lead times $\Delta t = 2$--48~h, the LSTM produces stable Dst forecasts and captures the major phases of geomagnetic activity, including storm-time main-phase depressions and recovery trends. The inclusion of cosmic-ray-derived features improves the model's sensitivity to impending CME-driven disturbances, particularly at intermediate and longer lead times where purely solar-wind-based predictors can be limited by propagation uncertainties and incomplete upstream coverage. Consistent with the physical picture of FD precursors, the model exhibits enhanced predictive skill during intervals characterized by coherent inter-station cosmic-ray suppression and elevated anomaly magnitudes. These results support the interpretation that cosmic-ray constraints provide an independent, physically grounded source of predictive information that complements standard solar-wind and geomagnetic indices in data-driven GST forecasting.

\subsection{Prediction Performance Evaluation}

\begin{figure}[ht!]
\plotone{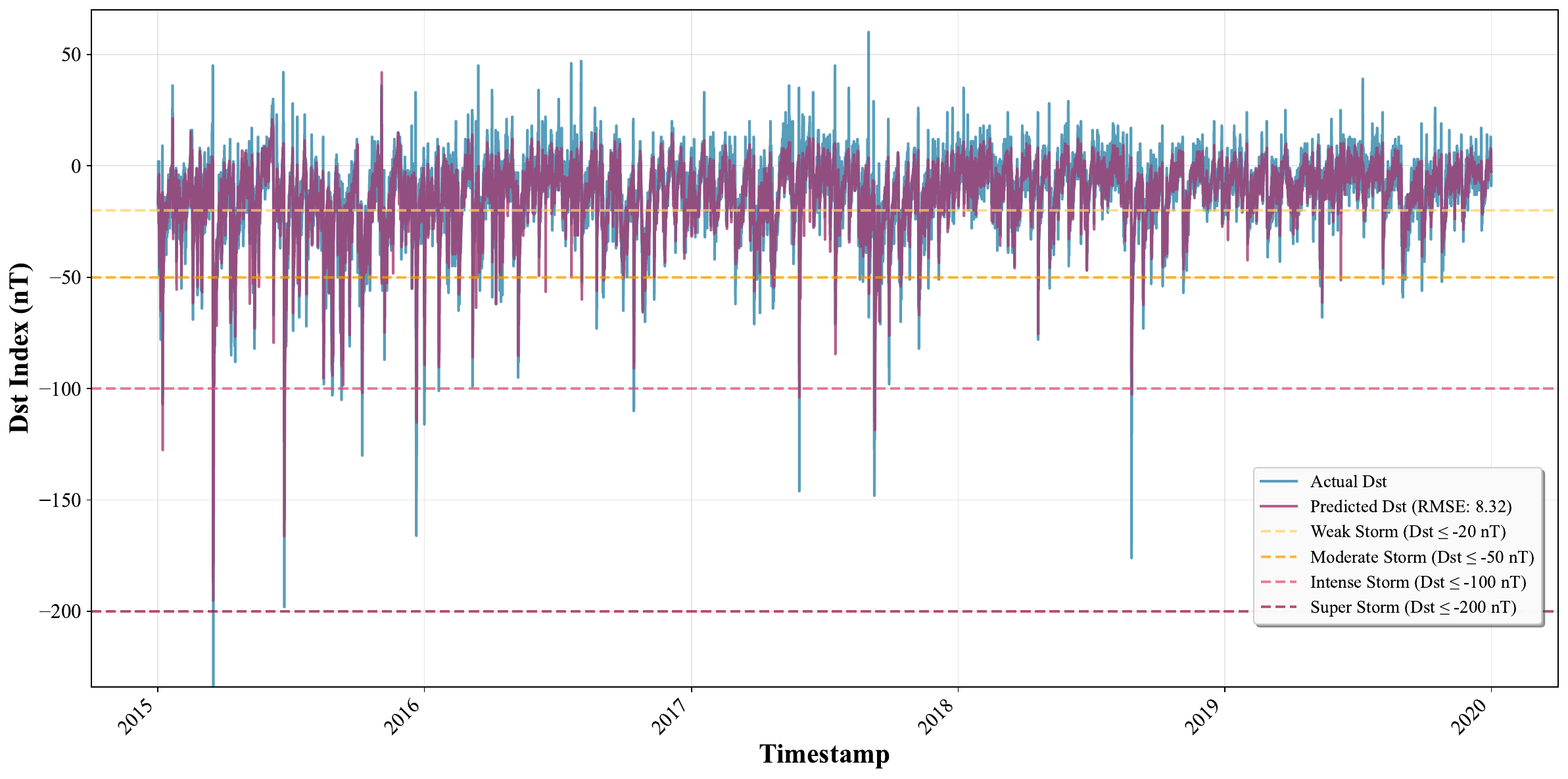}
\caption{Comparison of the 6~h-ahead Dst forecasts with observations for the test period.}
\label{fig:6h}
\end{figure}

Model performance is evaluated on both continuous Dst regression skill and storm-relevant classification skill. The full dataset is split into training and test subsets with a ratio of 0.8:0.2. The regression accuracy is quantified using the root-mean-square error (RMSE), which measures the overall deviation between predicted and observed Dst values:
\begin{equation}
\mathrm{RMSE}=\sqrt{\frac{1}{n}\sum_{i=1}^{n}\left(y_i-\hat{y}_i\right)^2},
\label{eq:rmse}
\end{equation}
where $y_i$ and $\hat{y}_i$ denote the observed and predicted Dst values at the $i$-th time step, respectively, and $n$ is the number of samples in the test set. The RMSE increases gradually with forecast lead time, reflecting the intrinsic predictability limit and accumulated uncertainty in upstream drivers: RMSE $=5.106$~nT (2~h), $8.315$~nT (6~h), $10.854$~nT (12~h), $12.883$~nT (24~h), and $14.788$~nT (48~h). Figure~\ref{fig:6h} provides a representative example for the 6~h forecasts, where the predicted Dst time series captures both the background variability and storm-time excursions with good phase agreement.

Because operational GST forecasting is primarily concerned with detecting storm occurrence and intensity, we further assess categorical prediction skill using precision, recall, and F1-score. For each storm class, we define true positives (TP), false positives (FP), and false negatives (FN) and compute
\begin{equation}
\mathrm{Precision}=\frac{\mathrm{TP}}{\mathrm{TP}+\mathrm{FP}},\qquad
\mathrm{Recall}=\frac{\mathrm{TP}}{\mathrm{TP}+\mathrm{FN}},\qquad
\mathrm{F1}=2\cdot\frac{\mathrm{Precision}\cdot\mathrm{Recall}}{\mathrm{Precision}+\mathrm{Recall}}.
\label{eq:f1}
\end{equation}
These metrics explicitly quantify the trade-off between missed storms (FN) and false alarms (FP), which is essential for practical warning applications.

We classify geomagnetic activity levels based on Dst thresholds:
\emph{Weak} ($-50~\mathrm{nT}<\mathrm{Dst}\leq -30~\mathrm{nT}$),
\emph{Moderate} ($-100~\mathrm{nT}<\mathrm{Dst}\leq -50~\mathrm{nT}$),
\emph{Intense} ($-200~\mathrm{nT}<\mathrm{Dst}\leq -100~\mathrm{nT}$), and
\emph{Super} ($\mathrm{Dst}\leq -200~\mathrm{nT}$).
For completeness, we also report the \emph{Quiet} category, corresponding to conditions outside the storm thresholds. This thresholding enables an evaluation of whether the model not only predicts Dst magnitudes accurately (regression) but also correctly identifies storm categories relevant to risk assessment.

We first perform a pointwise evaluation at the native 1-hour cadence by assigning each hourly Dst value (observed and predicted) to one of the above categories. Table~\ref{tab:6h_thresh} summarizes the 6~h-ahead classification statistics for all time points in the test set. The model achieves high skill in quiet-time identification (F1 $\approx 0.93$), while performance degrades with increasing storm intensity, consistent with the class-imbalance problem (rare extreme events) and the larger dynamical variability during strong storms. Nevertheless, the model maintains competitive precision for intense storms (0.828), indicating a relatively low false-alarm rate when it predicts a strong event, although recall is lower (0.438), implying that some intense storm hours are missed.

\begin{deluxetable}{lccccc}
\tablewidth{0pt}
\tablecaption{6-Hour Statistics and Predictions for Time Points Below Thresholds\label{tab:6h_thresh}}
\tablehead{
\colhead{Classification} & \colhead{Sample Count} & \colhead{Correct Predictions} &
\colhead{Precision} & \colhead{Recall} & \colhead{F1-score}
}
\startdata
Quiet & 32,912 & 31,049 & 0.914 & 0.943 & 0.928 \\
Weak & 9,621 & 6,589 & 0.737 & 0.685 & 0.710 \\
Moderate & 1,162 & 600 & 0.708 & 0.516 & 0.597 \\
Intense & 121 & 53 & 0.828 & 0.438 & 0.573 \\
Super & 3 & 0 & 0.00 & 0.00 & 0.00 \\
\enddata
\tablecomments{Precision = $\frac{\text{TP}}{\text{TP}+\text{FP}}$, Recall = $\frac{\text{TP}}{\text{TP}+\text{FN}}$, F1-score = $2 \times \frac{\text{Precision} \times \text{Recall}}{\text{Precision} + \text{Recall}}$, where TP, FP, and FN denote true positives, false positives, and false negatives, respectively.}
\end{deluxetable}

Pointwise scoring is sensitive to small phase errors and does not directly reflect operational requirements, where issuing a correct warning for a storm interval is often more important than matching every hourly value. We therefore introduce an event-based evaluation. A storm event is recorded when Dst remains within a given threshold range for at least 6 consecutive hours. For each observed event, a predicted event is considered successful if the temporal overlap between the predicted and observed intervals is at least 30\%. This overlap criterion allows for modest timing uncertainties while still requiring the model to capture the occurrence and approximate duration of storm intervals.

Table~\ref{tab:6h_event} reports event-based statistics for the 6~h forecasts. Compared to pointwise scoring, event-based precision and recall better reflect the model's ability to detect coherent storm episodes. The overall event-level F1-score reaches 0.533 for the aggregated storm set, with higher skill for moderate and intense storms than for weak storms. The absence of successful predictions for the super-storm category is expected given the extremely limited number of such events in the test set (only one event), underscoring the need for either longer archives, targeted sampling strategies, or physics-informed constraints specifically designed for rare extremes.

\begin{deluxetable}{lccccc}
\tablewidth{0pt}
\tablecaption{6-Hour Statistics and Predictions for GSTs Events\label{tab:6h_event}}
\tablehead{
\colhead{Classification} & \colhead{Sample Count} & \colhead{Correct Predictions} &
\colhead{Precision} & \colhead{Recall} & \colhead{F1-score}
}
\startdata
Weak & 253 & 110 & 0.588 & 0.435 & 0.500 \\
Moderate & 80 & 41 & 0.788 & 0.512 & 0.621 \\
Intense & 12 & 7 & 0.875 & 0.583 & 0.700 \\
Super & 1 & 0 & 0.00 & 0.00 & 0.00 \\
\hline
Overall & 346 & 158 & 0.640 & 0.457 & 0.533 \\
\enddata
\tablecomments{Storm events are defined as periods where Dst remains within a threshold range for at least 6 consecutive hours. A successful prediction requires $\geq$30\% temporal overlap between predicted and actual event intervals.}
\end{deluxetable}

Combining RMSE and storm-class skill provides a comprehensive view of model capability across forecasting horizons. RMSE increases monotonically from 2~h to 48~h, indicating progressively larger uncertainties at longer lead times, while categorical performance is primarily limited by (i) class imbalance for intense and super storms and (ii) timing errors that are more pronounced at larger $\Delta t$. The 6~h case (Figure~\ref{fig:6h} and Tables~\ref{tab:6h_thresh}--\ref{tab:6h_event}) demonstrates that the model can reproduce storm-time Dst depressions and achieve meaningful event-level detection skill, supporting its applicability to operationally relevant GST forecasting.

\subsection{Advantages of the Cosmic-Ray-Constrained Model}

A key motivation of this work is that cosmic-ray variations provide an upstream, physically grounded proxy of heliospheric disturbances that is complementary to in-situ solar-wind measurements. At longer forecast horizons, uncertainties in interplanetary propagation and incomplete upstream sampling can degrade purely solar-wind-driven predictors. By contrast, neutron monitor observations respond to the large-scale modulation produced by CME-driven shocks and magnetic clouds as they traverse the heliosphere, thereby offering additional information content that can persist at lead times beyond typical near-Earth solar-wind monitoring. This motivates the introduction of cosmic-ray modulation features as explicit constraints in the LSTM framework.

The advantage of incorporating cosmic-ray constraints becomes most evident for the 48-hour forecasting configuration. When the 48-hour model is trained \emph{without} cosmic-ray modulation features, the test performance yields $\mathrm{RMSE}=15.313$~nT and $\mathrm{F1}=0.178$. After adding the cosmic-ray constraints (pressure-corrected OULU/JUNG flux, 24-hour background levels, anomalies, and inter-station relative differences), the performance improves to $\mathrm{RMSE}=14.788$~nT and $\mathrm{F1}=0.224$. The F1-score gain corresponds to a relative improvement of 25.84\%, demonstrating that cosmic-ray information increases the model's capability to detect storm-category events at a 48-hour horizon. The RMSE reduction, while modest in absolute magnitude, is consistent with improved representation of storm-time Dst depressions that dominate the tail of the error distribution.

\begin{figure}[ht!]
\plottwo{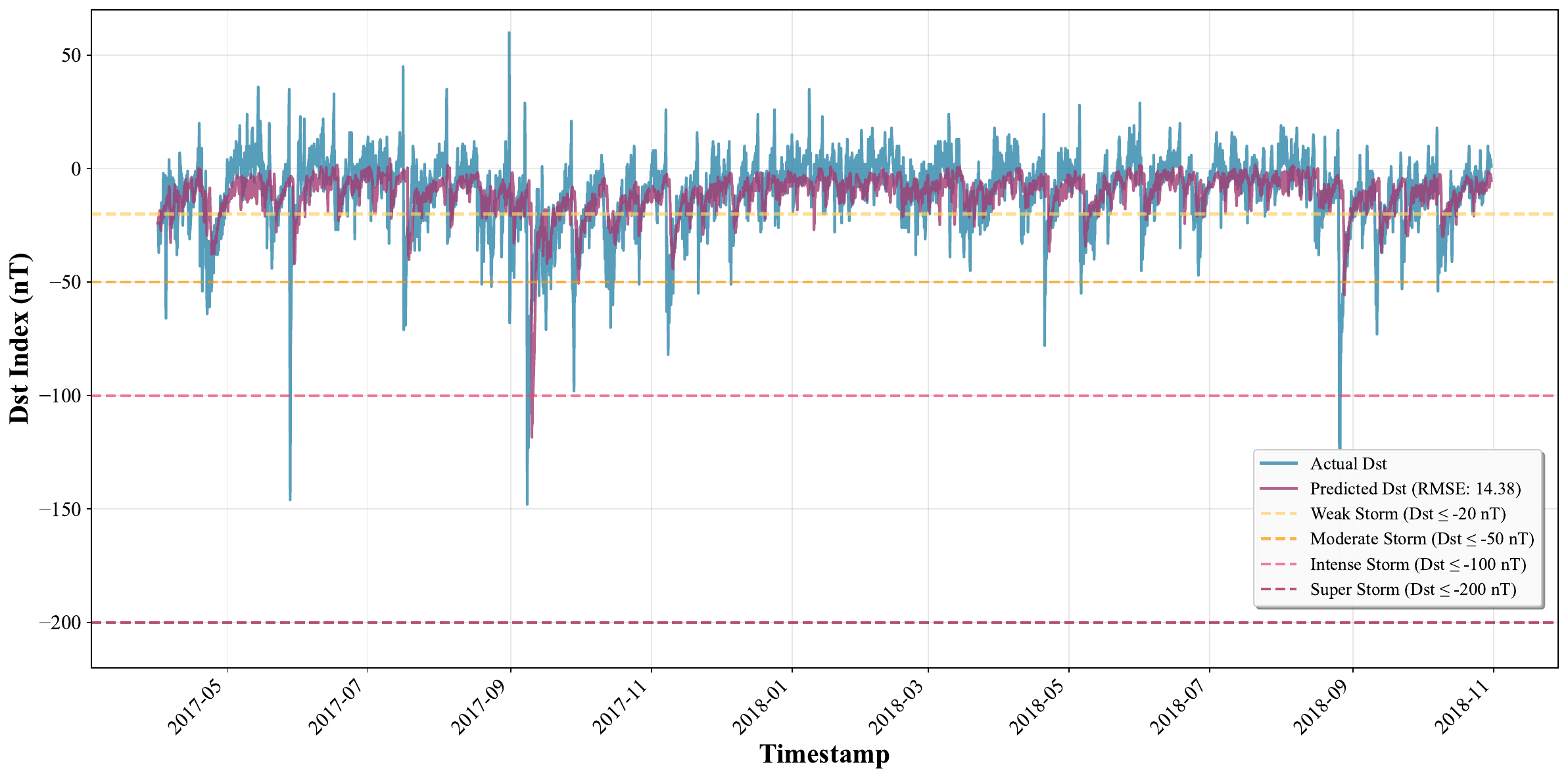}{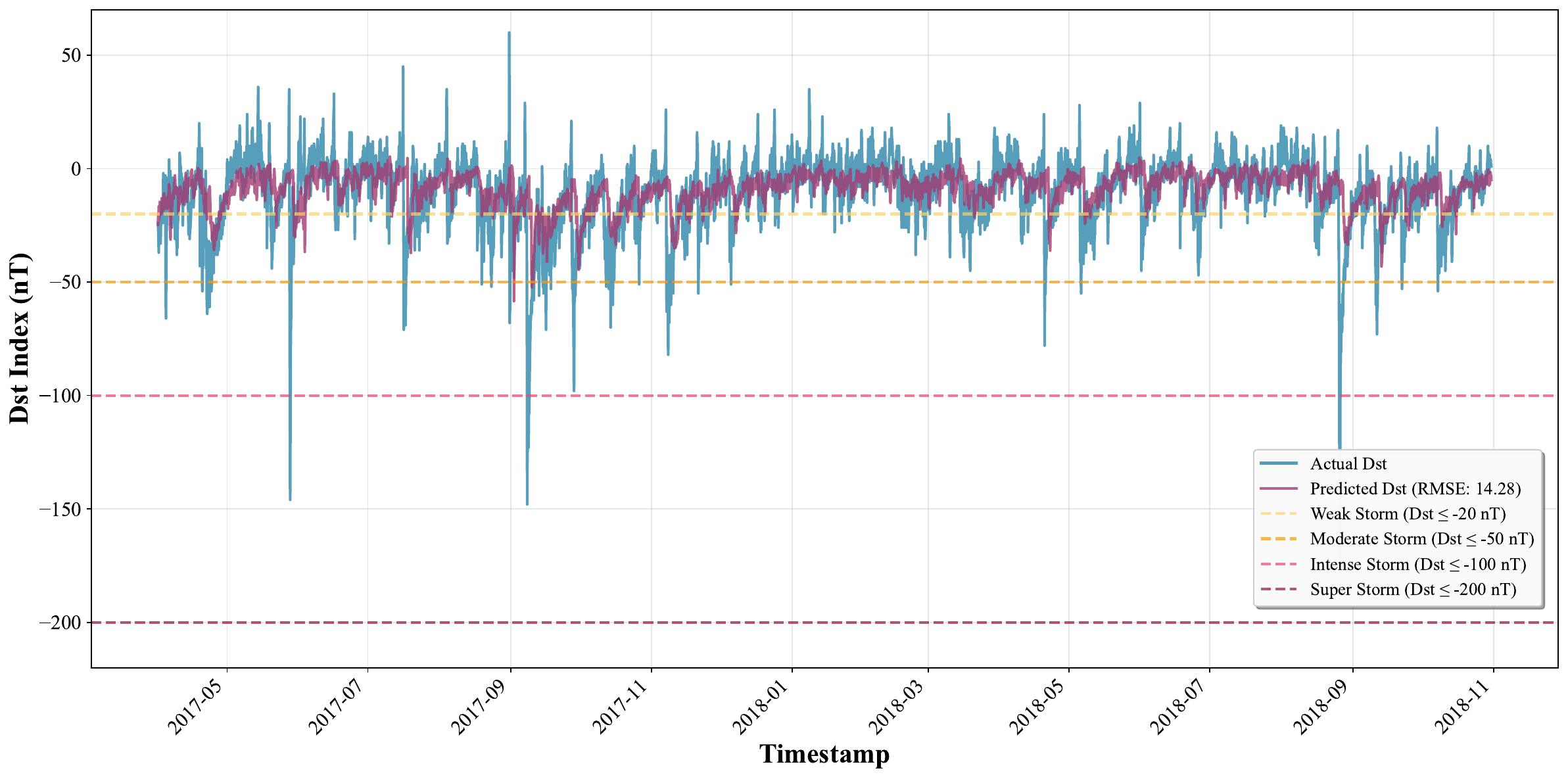}
\caption{48 hr Dst index forecasts for 2017--2018 with and without cosmic-ray modulation features.
         \textit{Left panel}: with cosmic-ray features. \textit{Right panel}: without cosmic-ray features.
         The model with cosmic-ray constraints exhibits improved performance during intense storm intervals.}
\label{fig:48h_comp}
\end{figure}

Figure~\ref{fig:48h_comp} compares 48-hour Dst forecasts for 2017--2018 produced by models trained with and without cosmic-ray features. The cosmic-ray-constrained model exhibits visibly improved tracking of intense storm intervals, with fewer missed events and reduced underestimation of the storm main-phase minimum. This behavior is physically plausible: strong CMEs often produce pronounced Forbush decreases and coherent inter-station cosmic-ray suppression, which are captured by the anomaly and relative-difference features. These signals provide early evidence of an approaching geoeffective disturbance and help the network distinguish CME-driven storm scenarios from quieter or CIR-dominated intervals.

From an operational perspective, the most valuable gain is the improvement in event-oriented skill at extended lead times, where traditional inputs may be insufficient. The enhanced F1-score indicates a better balance between missed storms and false alarms when cosmic-ray constraints are included. Because neutron monitor data are continuously available and robust, the proposed approach provides a low-cost and reliable augmentation to existing solar-wind-based predictors, particularly for 24--48~h lead-time warning applications that demand both timeliness and stability.

\subsection{Model Optimization}

\begin{figure}[ht!]
\plotone{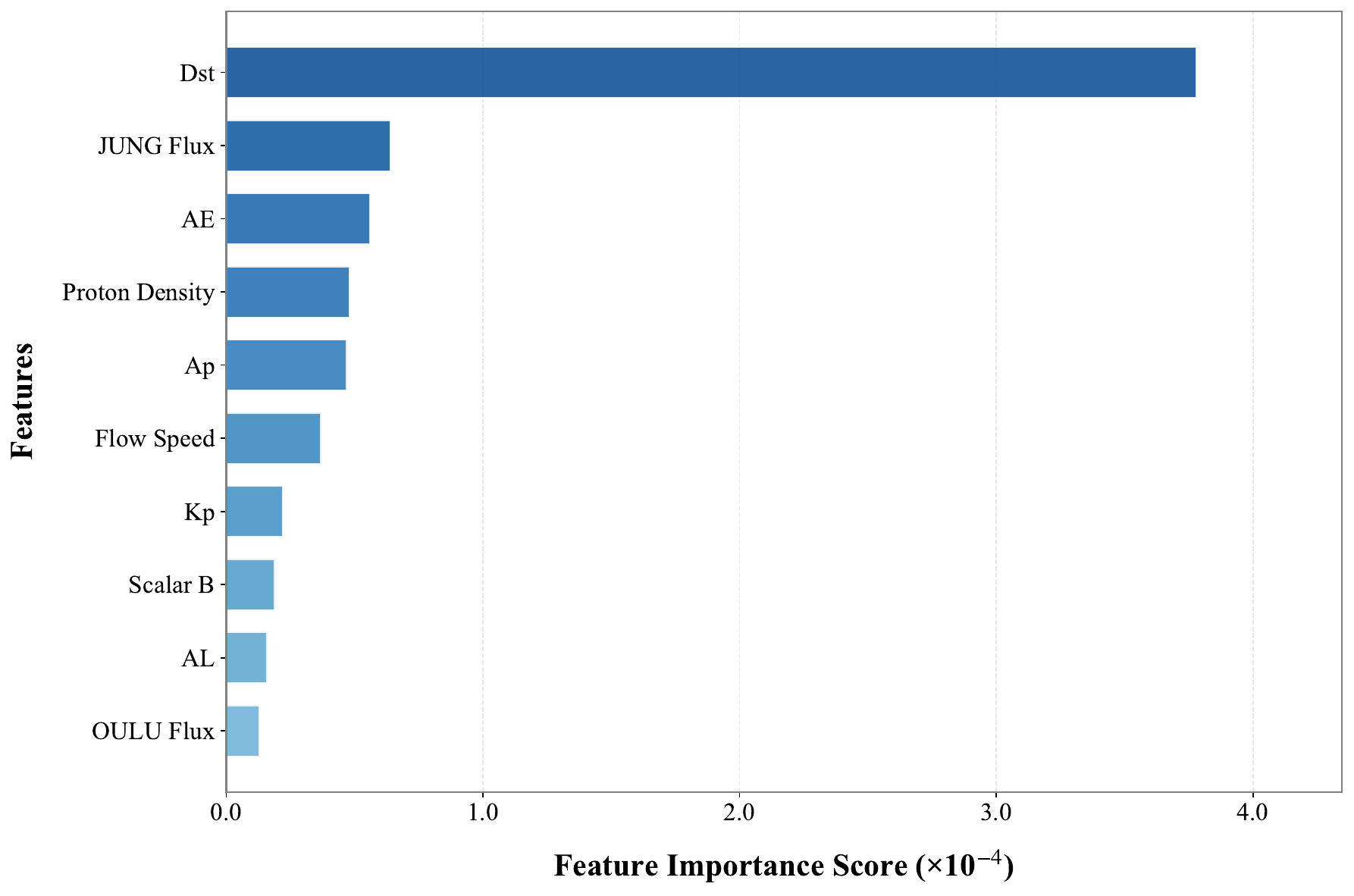}
\caption{Permutation Importance of input features for Dst Prediction}
\label{fig:importance}
\end{figure}

\begin{figure}[ht!]
\plotone{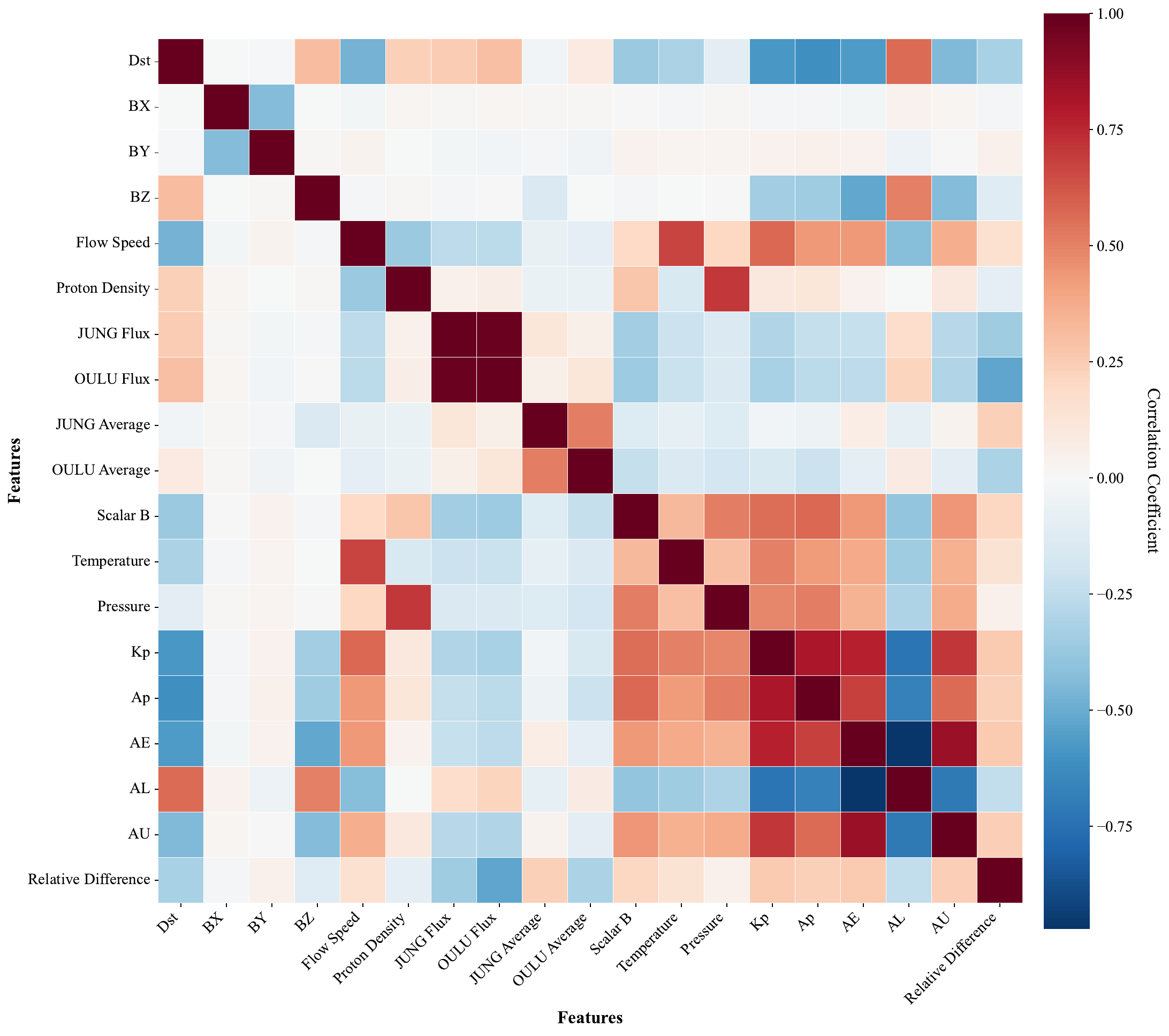}
\caption{Feature Correlation Heatmap used to diagnose multicollinearity in the input space.}
\label{fig:correlation}
\end{figure}

To improve model robustness and generalization---particularly for long-lead forecasting---we performed a targeted optimization that combines feature selection with hyperparameter tuning. The optimization is guided by two complementary diagnostics: permutation feature importance, which quantifies each feature's marginal contribution to predictive skill (Figure~\ref{fig:importance}), and pairwise correlation analysis, which identifies redundancy and multicollinearity within the input space (Figure~\ref{fig:correlation}). Together, these analyses enable us to remove low-value predictors and reduce correlated feature clusters without sacrificing physically meaningful information.

We first compute permutation importance by randomly shuffling each feature in the test set (or validation set) while keeping other inputs unchanged, and then measuring the degradation in prediction skill. Features that yield negligible performance change are considered weak contributors. Figure~\ref{fig:importance} displays the top 10 features ranked by their importance scores. Based on the complete importance analysis, IMF $B_X$ (GSE/GSM coordinate system) exhibits the lowest importance (importance $<0.002$) among all input features and is therefore removed from the feature set.

Next, we address multicollinearity using the correlation matrix (Figure~\ref{fig:correlation}). Severe redundancy can inflate the effective dimensionality, increase training variance, and reduce interpretability. In the auroral electrojet family, AL shows an extremely strong anticorrelation with AE ($r=-0.97$), and AU is strongly correlated with AE ($r=0.86$). Because AE already captures the overall auroral electrojet intensity and dominates predictive contribution among these indices, we exclude AL and AU while retaining AE to reduce redundancy and stabilize optimization.

Notably, although the 24-hour moving-average cosmic-ray backgrounds from Jungfraujoch and Oulu are highly correlated ($r=0.98$), we retain both features. This decision is supported by their strong individual permutation importance and by the fact that they originate from geographically separated stations with different cutoff rigidities and local responses. Retaining both preserves subtle station-dependent information that can be informative for distinguishing globally coherent CME-driven modulation from more structured variations.

After importance-driven pruning and correlation-informed redundancy reduction, the input feature space is reduced from 19 to 16 dimensions, improving parsimony while maintaining the physically meaningful drivers and cosmic-ray constraints.

In parallel, we optimize key model hyperparameters for the 48-hour prediction task. In particular, we conduct a grid-search over historical sequence length $L$ to determine the most informative look-back window for extended-lead forecasts. The best performance is achieved with $L=72$~hours, indicating that multi-day preconditioning of the solar wind, geomagnetic state, and cosmic-ray modulation provides measurable predictive value for the 48-hour horizon.

The refined 16-feature model with an optimized 72-hour historical window yields improved test performance relative to the baseline cosmic-ray-constrained configuration. Specifically, RMSE decreases from 14.788~nT to 14.695~nT. Event-oriented skill also improves: overall precision increases from 0.335 to 0.341, recall from 0.168 to 0.179, and F1-score from 0.224 to 0.235. These gains confirm that targeted feature selection and hyperparameter tuning can enhance long-lead geomagnetic storm prediction by reducing redundant inputs, improving conditioning, and better matching the temporal memory of the LSTM to the forecasting horizon.

\subsection{Model Validation}
\label{sec:validation}

To assess the robustness and generalization of the proposed LSTM framework, we conducted two complementary validation experiments for the 48-hour forecast configuration: (i) five-fold cross-validation with temporally ordered splits and (ii) an ablation study on LSTM depth. Together, these tests evaluate stability against sampling variability and verify that the selected model complexity is appropriate for the available data, thereby reducing the risk of overfitting \citep{kohavi1995study}.

\begin{figure}[ht!]
\plotone{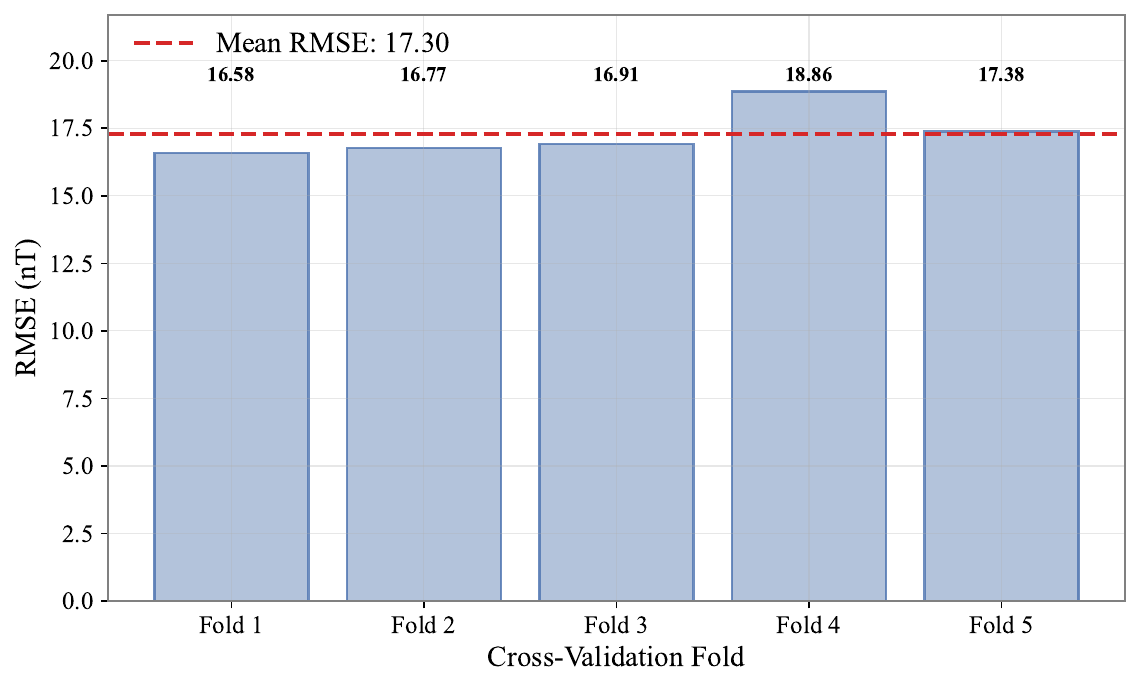}
\caption{LSTM model RMSE across five cross-validation folds (48-hour horizon).}
\label{fig:cv_rmse}
\end{figure}

We performed five-fold cross-validation on the full 1995--2020 hourly dataset while preserving temporal order. The series was partitioned into five contiguous subsets of equal length. In each fold, one subset was used as the validation set and the remaining four subsets were used for training, ensuring that the model is repeatedly evaluated on unseen intervals under a consistent chronological splitting strategy. Figure~\ref{fig:cv_rmse} summarizes the validation RMSE across folds.

The RMSE values for the five folds are 16.58~nT, 16.77~nT, 16.91~nT, 18.86~nT, and 17.38~nT. The mean RMSE is 17.30~nT with a standard deviation of 0.87~nT, indicating that performance is consistent across different temporal segments and that the model does not rely on a specific sub-interval for skill. The fourth fold exhibits the largest RMSE (18.86~nT), suggesting that this period likely contains more challenging storm-time variability; nevertheless, the spread remains limited, supporting overall robustness.

\begin{figure}[ht!]
\plotone{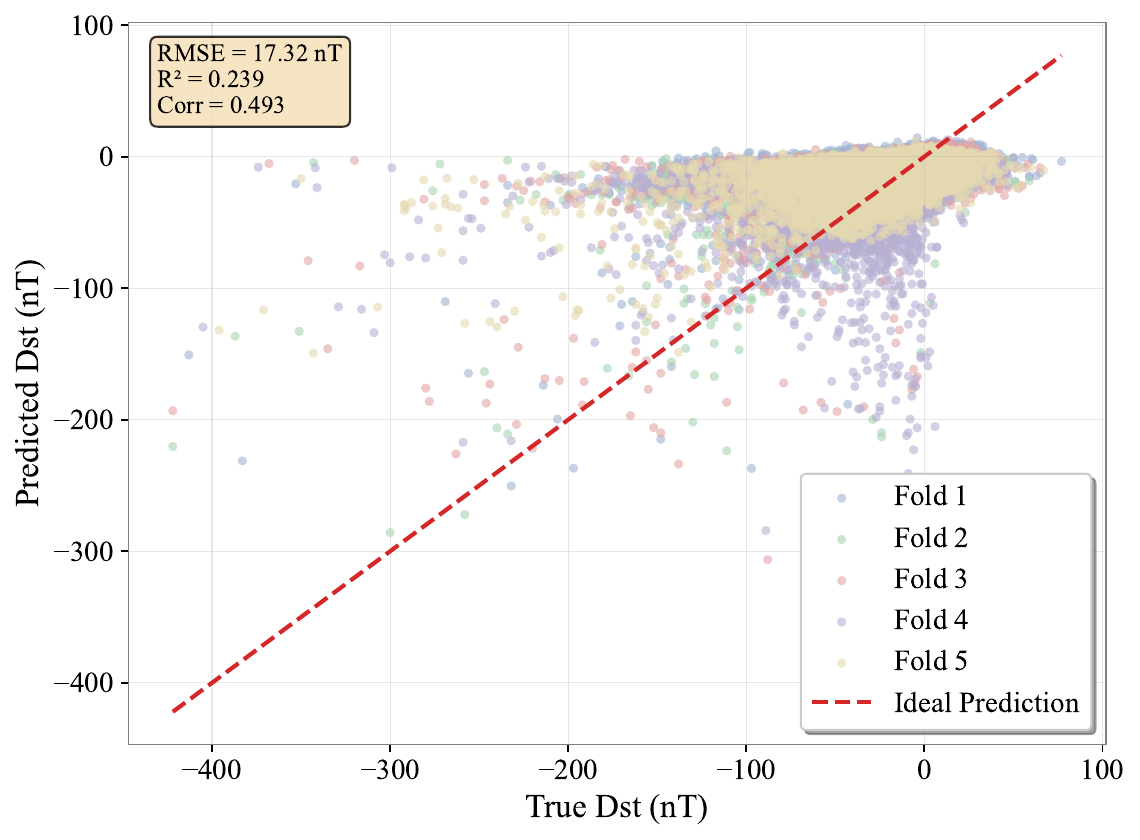}
\caption{Predicted versus observed Dst values across cross-validation folds, color-coded by fold.}
\label{fig:cv_scatter}
\end{figure}

To further evaluate agreement between predictions and observations, Figure~\ref{fig:cv_scatter} presents predicted versus observed Dst values aggregated over all folds and color-coded by fold. The overall statistics show RMSE $\approx 17.32$~nT, $R^2=0.239$, and a correlation coefficient of 0.493, indicating meaningful linear association at a 48-hour horizon while leaving substantial variance unexplained---a reflection of the intrinsic predictability limits and the complexity of storm-time dynamics.

\begin{figure}[ht!]
\plotone{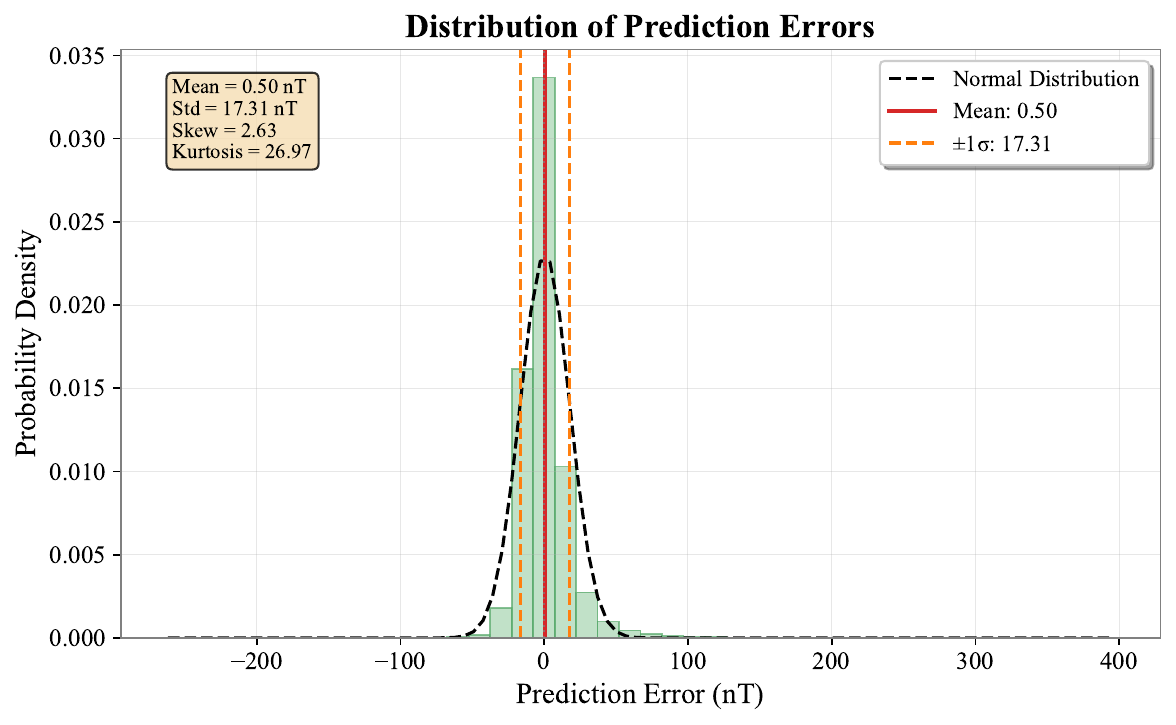}
\caption{Distribution of prediction errors across cross-validation folds, overlaid with a fitted normal distribution.}
\label{fig:cv_error}
\end{figure}

Residual behavior is examined in Figure~\ref{fig:cv_error}, which shows the distribution of prediction errors pooled across folds. The error distribution is centered near zero (mean $\approx 0.50$~nT) with a standard deviation of 17.31~nT, indicating that the model is approximately unbiased in the mean sense. The distribution exhibits pronounced non-Gaussian tails (skewness 2.63; kurtosis 26.97), which is expected because large Dst excursions are rare and difficult to predict at long lead times. These heavy tails imply that extreme storm-time deviations dominate the error budget and motivate the use of event-based and categorical metrics in addition to RMSE when evaluating operational performance.

\begin{figure}[ht!]
\centering
\plottwo{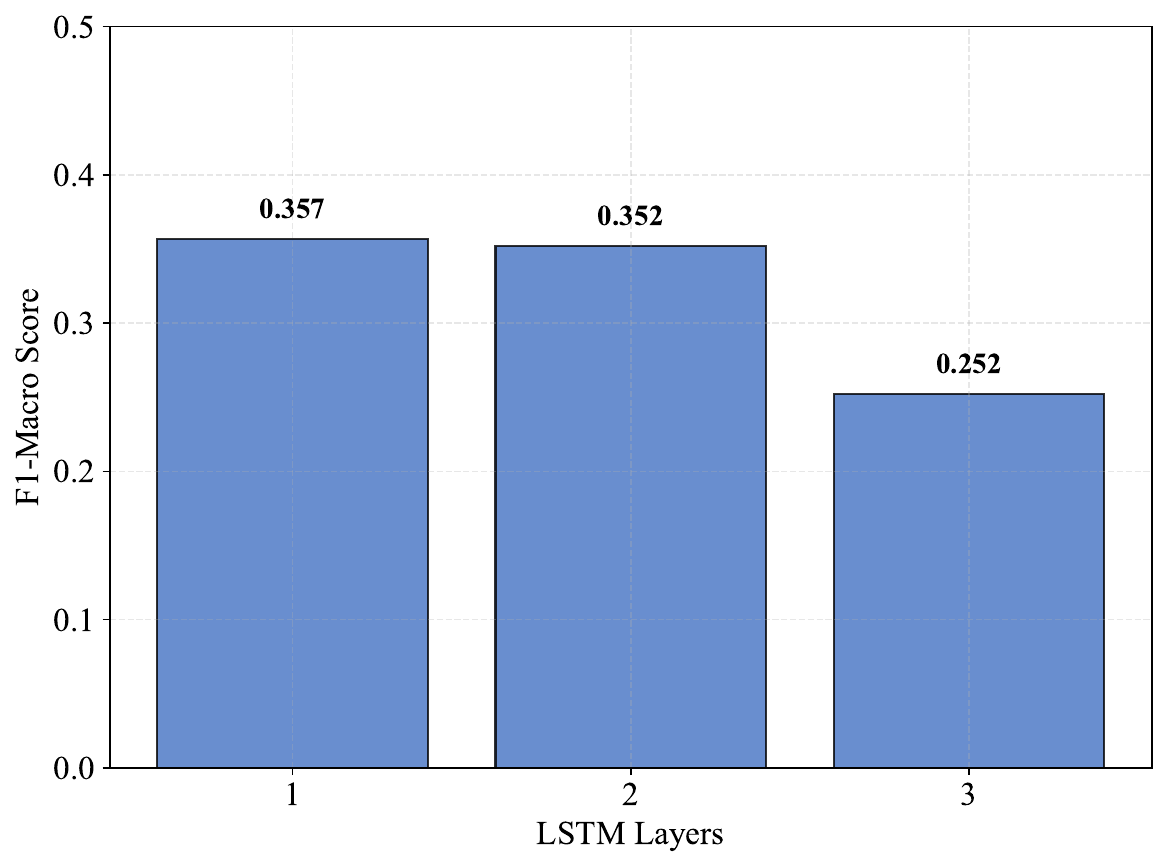}{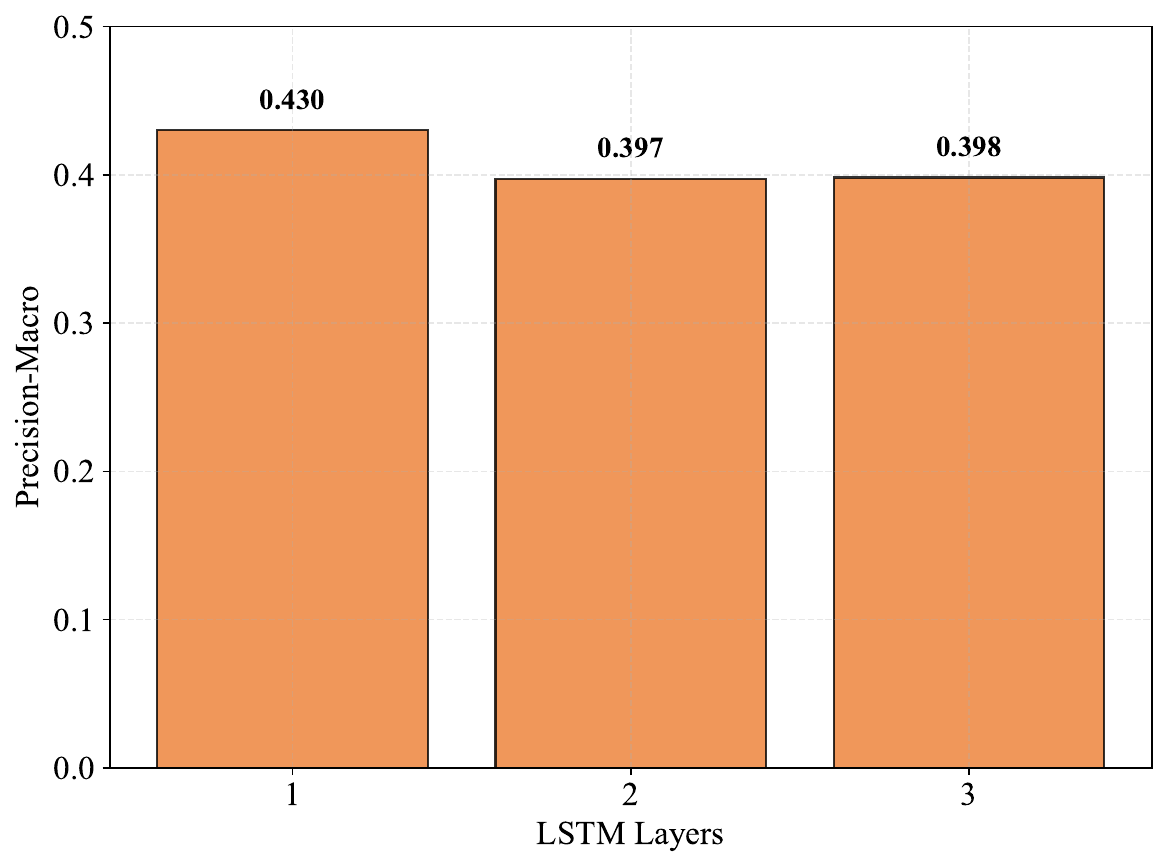}
\vspace{10pt}
\plottwo{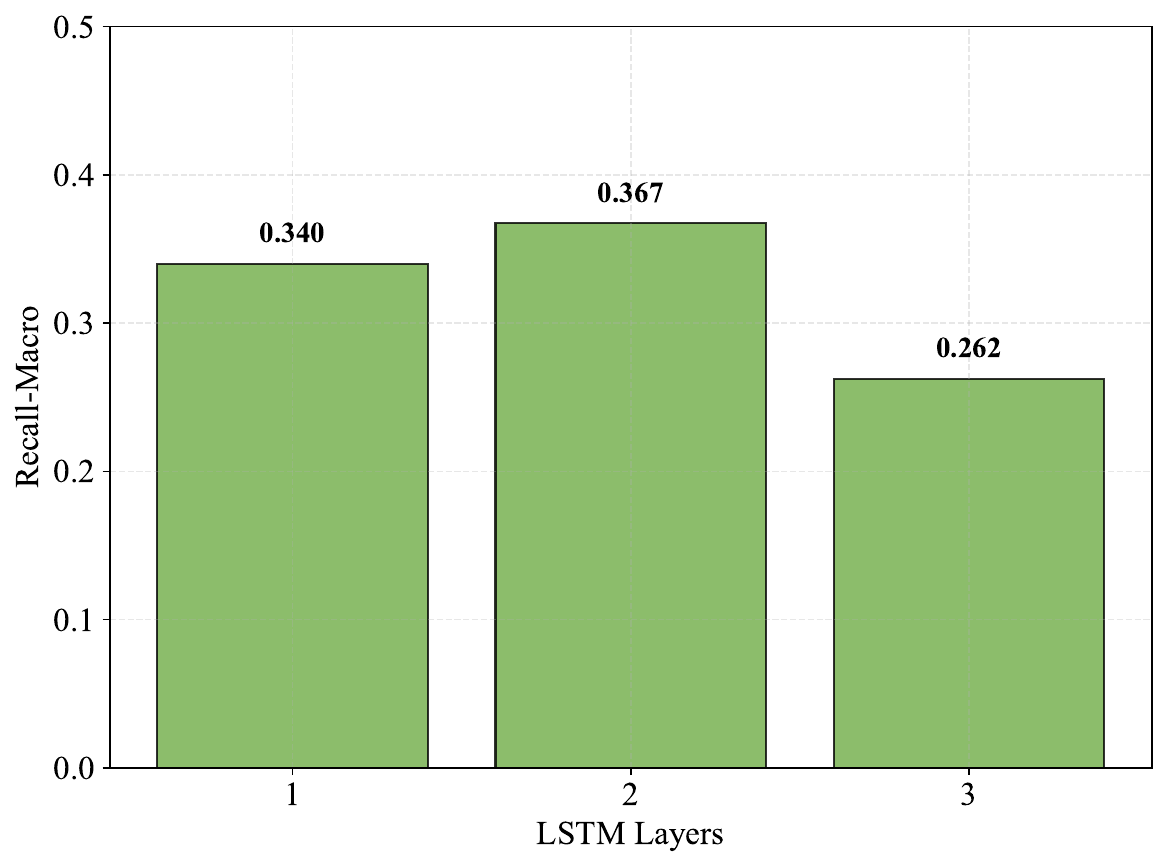}{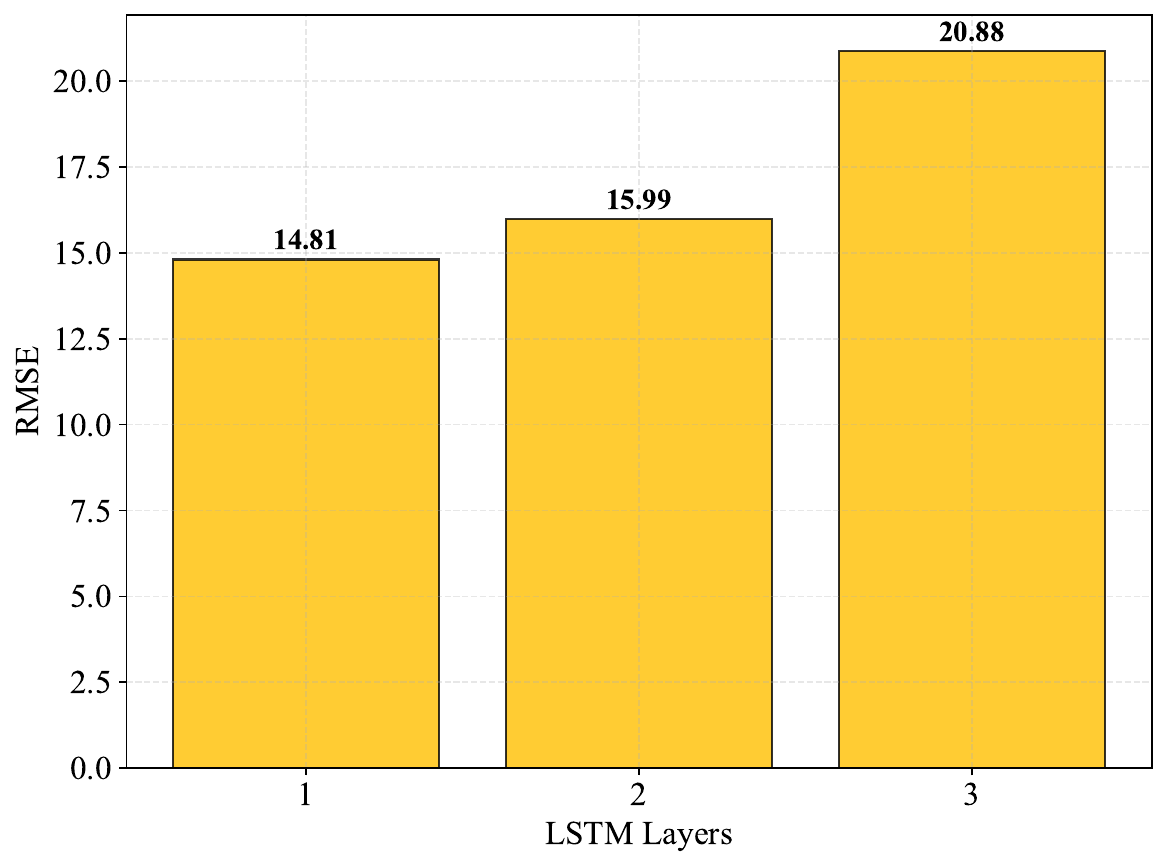}
\caption{Performance metrics versus number of LSTM layers (48 hr forecast horizon).
         \textit{Top left}: F1-macro score; \textit{Top right}: precision-macro;
         \textit{bottom left}: recall-macro; \textit{bottom right}: RMSE.
         The single-layer model achieves the best overall performance.}
\label{fig:ablation_f1}
\end{figure}

In addition to cross-validation, we tested the sensitivity of performance to model depth by training LSTM architectures with one, two, and three recurrent layers under the same 48-hour forecasting setup. To isolate the effect of depth, each recurrent layer contains 50 hidden units (excluding input and output layers), and training uses identical data, preprocessing, and evaluation procedures. We report macro-averaged precision, recall, and F1-score---which treat each storm category equally---together with RMSE. Figure~\ref{fig:ablation_f1} summarizes these metrics as a function of depth.

The single-layer configuration achieves the best overall balance between accuracy and complexity, with F1-macro = 0.357, precision-macro = 0.430, recall-macro = 0.340, and RMSE = 14.81~nT. Increasing depth to two layers produces a modest increase in recall-macro (0.367) but reduces precision-macro (0.397) and F1-macro (0.352), and degrades RMSE to 15.99~nT while substantially increasing computational cost. The three-layer model performs markedly worse across all metrics (F1-macro = 0.252; precision-macro = 0.398; recall-macro = 0.262; RMSE = 20.88~nT), indicating over-parameterization and overfitting for this task.

The temporally ordered five-fold cross-validation demonstrates stable predictive skill across different periods of the 1995--2020 record, with limited RMSE variability and near-zero mean bias. The depth ablation further confirms that a single-layer LSTM with 50 units provides the most reliable performance, whereas deeper architectures introduce unnecessary complexity and degrade generalization. These results support the robustness of the proposed cosmic-ray-constrained LSTM model and justify the selected architecture for long-lead geomagnetic storm forecasting.

\subsection{LSTM Neural Network Architecture}

We employ a long short-term memory (LSTM) neural network as the core predictor for Dst forecasting. Compared with conventional recurrent neural networks (RNNs), LSTM introduces an explicit memory cell regulated by gating operations, which alleviates the vanishing-gradient problem and enables learning of long-range temporal dependencies \citep{Gers2000,greff2016lstm}. This capability is particularly important for geomagnetic storms, whose evolution is intrinsically multi-phase (initial, main, and recovery) and depends on both short-lived solar-wind triggers and longer-term preconditioning of the magnetosphere.

At each time step $t$, the network takes as input a multivariate feature vector $\mathbf{x}_t$ and updates its hidden state $\mathbf{h}_t$ and cell state $\mathbf{c}_t$ through three gates. The forget gate $\mathbf{f}_t$ determines how much past memory is retained, the input gate $\mathbf{i}_t$ controls the amount of new information written into memory, and the output gate $\mathbf{o}_t$ regulates how much of the updated cell state is exposed to the hidden representation. These operations can be expressed as
\begin{equation}
\mathbf{f}_t=\sigma\!\left(\mathbf{W}_f[\mathbf{h}_{t-1},\mathbf{x}_t]+\mathbf{b}_f\right),\quad
\mathbf{i}_t=\sigma\!\left(\mathbf{W}_i[\mathbf{h}_{t-1},\mathbf{x}_t]+\mathbf{b}_i\right),\quad
\mathbf{o}_t=\sigma\!\left(\mathbf{W}_o[\mathbf{h}_{t-1},\mathbf{x}_t]+\mathbf{b}_o\right),
\end{equation}
where $\sigma(\cdot)$ denotes the sigmoid activation, $[\cdot,\cdot]$ represents concatenation, and $\mathbf{W}_*$ and $\mathbf{b}_*$ are trainable weights and biases. A candidate memory update $\tilde{\mathbf{c}}_t$ is computed and combined with the previous cell state to form the new cell state:
\begin{equation}
\tilde{\mathbf{c}}_t=\tanh\!\left(\mathbf{W}_c[\mathbf{h}_{t-1},\mathbf{x}_t]+\mathbf{b}_c\right),\qquad
\mathbf{c}_t=\mathbf{f}_t\odot \mathbf{c}_{t-1} + \mathbf{i}_t\odot \tilde{\mathbf{c}}_t,
\end{equation}
followed by the hidden-state update
\begin{equation}
\mathbf{h}_t=\mathbf{o}_t\odot \tanh(\mathbf{c}_t),
\end{equation}
where $\odot$ denotes element-wise multiplication. Through these gated updates, the model can adaptively preserve information over long intervals while remaining responsive to abrupt changes in the solar wind and cosmic-ray precursors, thereby matching the multi-timescale nature of GST evolution.

The adopted architecture consists of a single LSTM layer with 50 hidden units, followed by a fully connected output layer that maps the final hidden representation (or an aggregated sequence representation) to the predicted Dst value at the specified lead time. This single-layer configuration is selected to balance predictive skill and generalization, consistent with the validation results reported in Section~\ref{sec:validation}. In implementation, a rectified linear unit (ReLU) activation is used in the subsequent dense mapping stage to enhance nonlinearity in the final regression, while the internal LSTM gating and state updates retain their standard sigmoid/tanh forms.

Model training minimizes the mean squared error (MSE) between predicted and observed Dst values. We use the Adam optimizer with an initial learning rate of 0.001, a batch size of 32, and standard backpropagation through time (BPTT) to update parameters. These settings provide stable convergence for long time-series sequences while maintaining computational efficiency. Unless otherwise noted, the same architecture and optimization configuration are used across all forecast horizons to enable a consistent comparison of predictive performance.

\subsection{Cosmic-Ray Modulation as Feature Input}

Cosmic-ray modulation is introduced in this work as a physics-informed feature engineering strategy to augment conventional solar-wind-driven predictors. GCRs propagate at nearly the speed of light ($\sim 3\times10^5$~km~s$^{-1}$) and therefore respond almost immediately (in an observational sense) to large-scale heliospheric disturbances as they evolve and sweep through interplanetary space. In contrast, the associated solar-wind structures that ultimately drive GSTs typically travel at a much slower speed, usually $\sim$300--800~km~s$^{-1}$. This fundamental disparity in propagation speed means that cosmic-ray signatures produced by an approaching CME-driven shock or ejecta can be observed at Earth by ground-based neutron monitors well in advance of the arrival of the slower solar-wind disturbance, providing a practical early-warning channel on the order of 1--3 days..

\begin{figure}[ht!]
\plotone{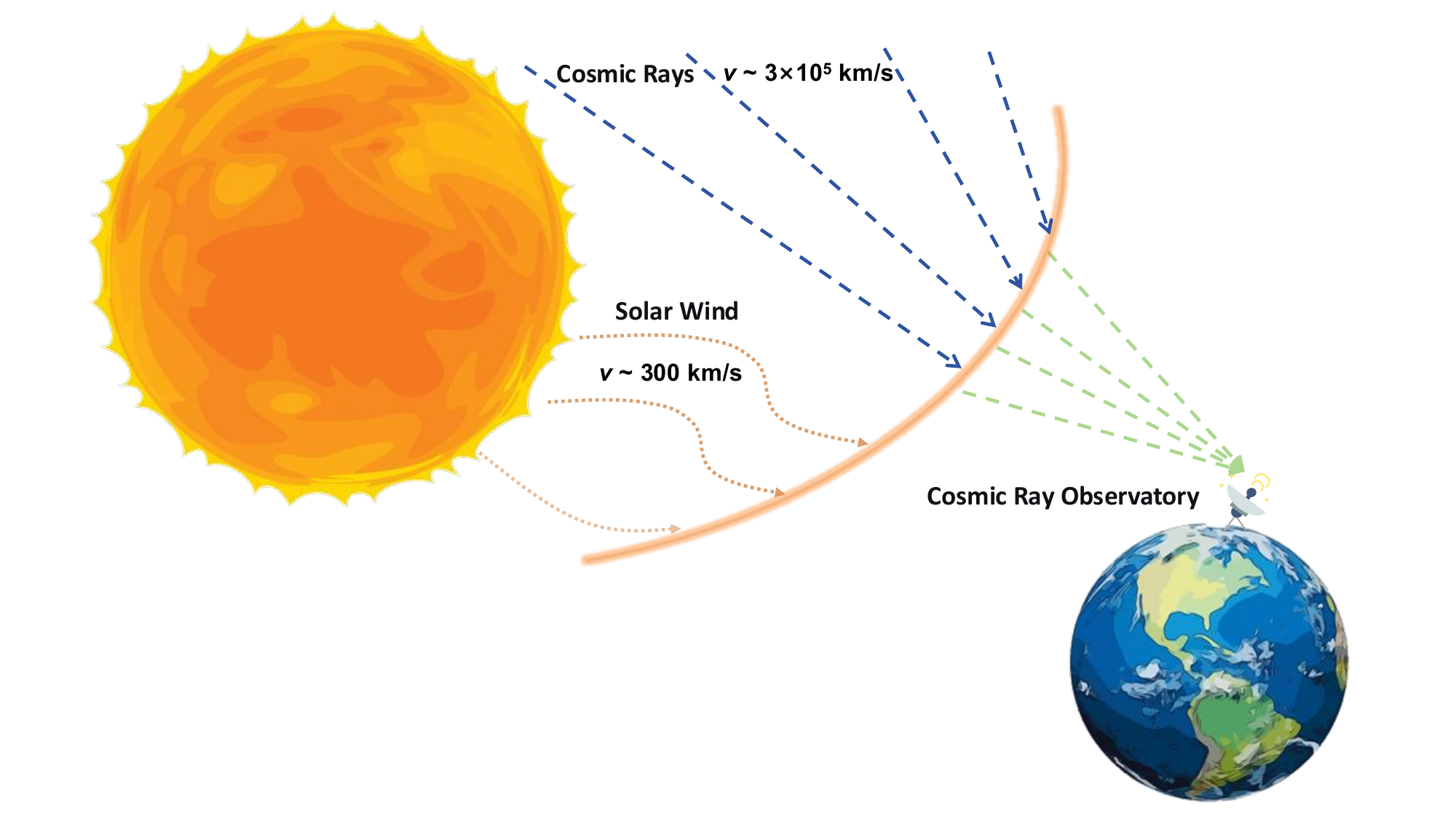}
\caption{Schematic illustration of the early-warning mechanism provided by GCRs. High-energy cosmic rays (blue/green arrows) are modulated by an approaching heliospheric disturbance and are observed at Earth-based neutron monitors ahead of the slower solar-wind structure (orange arrows) that later triggers the geomagnetic storm.}
\label{fig:precursor}
\end{figure}

When a CME and/or its driven shock propagates through the heliosphere, the enhanced IMF strength, turbulence, and compressed plasma in the disturbance increase cosmic-ray scattering and effectively reduce the GCR flux reaching the inner heliosphere. The resulting decrease in neutron monitor count rates is commonly referred to as a FD. The onset and depth of the FD encode information about the strength and spatial extent of the approaching disturbance, and thus can serve as a proxy for impending geoeffective conditions. Importantly, because neutron monitors sample the integrated modulation of energetic particles, they provide a complementary diagnostic to local in-situ measurements, which can be limited by spacecraft coverage and propagation uncertainty at long lead times.

We incorporate neutron monitor observations from two geographically separated stations---Oulu (OULU) and Jungfraujoch (JUNG)---to construct three categories of predictive features that represent both the absolute modulation and its spatial coherence.

\noindent\textit{(1) Background level and raw flux.}
We use the pressure-corrected hourly flux at each station as the primary observable and compute a 24-hour moving average to represent the slowly varying background level. The moving average is constructed with a one-hour lag window, such that the statistic at time $t$ uses only past observations (from $t-24$ to $t-1$), which prevents data leakage in forecasting experiments.

\noindent\textit{(2) FD-related transient disturbance indicators.}
To isolate short-term modulation associated with heliospheric transients, we compute FD-like indicators as deviations from the background level, i.e., the difference between the instantaneous flux and the 24-hour moving average. These anomaly features emphasize rapid depressions and recoveries that are not captured by the background trend alone and are particularly relevant for CME-driven disturbances.

\noindent\textit{(3) Inter-station coherence features.}
Single-station measurements can be affected by local effects and may not fully capture the global coherence of heliospheric disturbances. We therefore derive inter-station features that quantify spatial consistency, including sliding-window Pearson correlation coefficients and normalized relative differences between OULU and JUNG. In particular, the relative-difference feature (Equation~\ref{eq:reldiff}) highlights differential station responses while suppressing common-mode variability, providing a robust descriptor of large-scale modulation patterns.

All cosmic-ray-derived features are Min--Max normalized and concatenated with the solar-wind/IMF and geomagnetic parameters to form the final 19-dimensional input vector (Table~\ref{tab:features}). The effectiveness of this physics-informed augmentation is confirmed by ablation experiments at the 48-hour horizon. Without cosmic-ray modulation features, the model achieves $\mathrm{RMSE}=15.313$~nT and $\mathrm{F1}=0.178$ for GST event detection. After incorporating the cosmic-ray constraints, performance improves to $\mathrm{RMSE}=14.788$~nT and $\mathrm{F1}=0.224$, corresponding to a 25.84\% relative increase in F1-score. These results demonstrate that cosmic-ray modulation provides independent and operationally valuable predictive information, particularly for extended-lead GST forecasting where reliance on in-situ solar-wind drivers alone can be insufficient.

\section{Discussion}

This study demonstrates that fusing cosmic-ray modulation signatures with an LSTM forecasting framework can improve geomagnetic storm (GST) prediction skill, particularly at extended lead times. The key implication is that neutron monitor observations provide an upstream, physically grounded information channel that complements near-Earth in-situ solar-wind measurements, thereby helping to mitigate the loss of predictability associated with interplanetary propagation uncertainty. Nevertheless, several methodological and data-driven limitations remain, and addressing them will be essential for improving robustness, interpretability, and operational readiness.

A central strength of the proposed approach is its physics-informed feature design: cosmic-ray background, anomaly, and inter-station coherence features are constructed to encode signatures of large-scale heliospheric disturbances (e.g., FD-like modulation) that often precede intense storms. However, the current framework still relies on a relatively simple model formulation. First, Dst is used as the sole prediction target. While Dst is a widely adopted index of ring-current intensity, it does not fully characterize storm-time dynamics across different current systems and latitudes. Exclusive reliance on Dst may therefore limit the model's ability to discriminate between storm drivers and to capture substorm-related or high-latitude responses.

Second, although the input space merges solar-wind, geomagnetic, and cosmic-ray information, the LSTM architecture remains relatively compact. This design is advantageous for stability and real-time deployment, but it may not fully exploit nonlinear interactions among multi-source drivers, especially during compound events where CME and CIR influences overlap. This limitation is reflected in the reduced recall for rare strong storms, where class imbalance and highly nonlinear coupling dominate the forecast uncertainty.

Third, cosmic-ray modulation is incorporated as feature input rather than as an explicit physical constraint. Under extreme conditions, purely data-driven mappings may extrapolate poorly, particularly if the training archive contains limited examples of super storms. Embedding additional physics into the learning process may therefore improve generalization and interpretability beyond what input-level constraints alone can offer.

One immediate extension is hybrid modeling that combines the strengths of sequential neural networks with robust nonlinear classifiers. For example, LSTM can be used as a temporal feature extractor, while an ensemble method such as random forests can operate on learned representations (and/or physically engineered features) to improve event-level classification, especially under imbalanced storm categories \citep{Breiman2001}. Such a two-stage design can also enhance interpretability through feature importance rankings and reduce sensitivity to noise in rare-event regimes.

A complementary direction is to incorporate explicit physical constraints using physics-informed machine learning. In the present work, cosmic-ray modulation enters as an empirical precursor, but the model does not enforce quantitative consistency between cosmic-ray signatures and magnetospheric energy input. Future work could establish data-driven but physically interpretable relationships linking cosmic-ray depletion or coherence measures to coupling functions (e.g., proxies related to reconnection or solar-wind energy transfer) and embed these relations as soft constraints or regularization terms. Physics-informed neural network paradigms provide a potential pathway to integrate such constraints and improve extrapolation behavior under extremes \citep{Raissi2019}.

Beyond their role as auxiliary predictors within an LSTM, cosmic rays can potentially serve as an independent early-warning channel. A pragmatic operational approach is to build a rule-based detector using FD-related indicators (e.g., anomaly magnitude, FD depth, recovery rate) and inter-station coherence metrics. For example, an alert could be triggered when FD depth exceeds a station-validated threshold and the OULU--JUNG coherence increases sharply within a sliding window, indicating a globally coherent heliospheric disturbance. Such a threshold-based discriminator would be simple, transparent, and computationally inexpensive, and could complement the neural-network forecasts by providing conservative ``watch'' signals at longer lead times. Importantly, thresholds should be calibrated using long-term statistics and conditioned on solar-cycle phase and background modulation levels to reduce false alarms.

An important limitation is that the current hourly dataset does not explicitly include CME physical parameters such as eruption strength, propagation direction, shock properties, and magnetic configuration (including the polarity and the likelihood of southward IMF at 1~AU). These attributes are directly linked to both cosmic-ray modulation and storm geoeffectiveness, and their omission constrains the model's physical fidelity. At present, the primary barrier is data availability: continuous and consistently calibrated CME parameters at 1-hour cadence are not routinely available over multi-decade intervals.

A promising future extension is to construct a harmonized database that aligns CME catalogs and remote-sensing diagnostics (e.g., speed, width, source longitude, inferred arrival time, and magnetic proxies where available) with solar-wind, cosmic-ray, and geomagnetic measurements on a common timeline. With such a dataset, supervised learning could explicitly condition forecasts on CME attributes, potentially improving robustness for extreme events and reducing ambiguities between CME-driven and CIR-driven storms.

This work points to several broader opportunities. First, extending the framework from single-target (Dst) prediction to multi-task learning could improve physical consistency by jointly predicting multiple geomagnetic indices (e.g., Dst, AE, Kp/Ap), thereby enabling the model to learn shared drivers and index-specific sensitivities. Second, incorporating uncertainty quantification (e.g., probabilistic forecasting or ensemble approaches) would better support operational decision-making by providing confidence bounds for storm warnings. Third, continued refinement of cosmic-ray modulation features---including station networks beyond two sites, rigidity-dependent responses, and disturbance morphology descriptors---may further strengthen the precursor signal.

Finally, we note that more speculative directions (e.g., quantum computing approaches for sequence learning) remain exploratory and are currently not required to realize tangible forecasting improvements. In the near term, the most impactful advances are likely to come from (i) improved storm-event handling under class imbalance, (ii) explicit physics constraints that link cosmic-ray signatures to coupling functions, and (iii) enriched CME-parameter datasets. Advancing along these directions will help develop more robust, interpretable, and operationally reliable space-weather forecasting models \citep{Camporeale2019}.

\section{Conclusions}

In this study, we develop a physics-informed LSTM framework for GST prediction by explicitly integrating cosmic-ray modulation signatures as precursor features. Using a 25-year multi-source space-environment dataset (1995--2020) with 1-hour cadence, the proposed model demonstrates that GCR variations measured by ground-based neutron monitors contain actionable early-warning information on the order of 2--3 days before major geomagnetic disturbances. This precursor channel complements conventional solar-wind and IMF drivers and provides additional predictive value, particularly for extended lead-time forecasts.

Quantitatively, the model achieves RMSE values ranging from 5.106 to 14.788~nT for forecast horizons of 2--48~hours. The benefit of cosmic-ray constraints is most pronounced at long lead times: for the 48-hour forecasting task, incorporating cosmic-ray modulation features improves event-detection skill, increasing the F1-score from 0.178 to 0.224 (a 25.84\% relative gain) and reducing RMSE from 15.313~nT to 14.788~nT compared with an otherwise identical LSTM trained without cosmic-ray inputs. These results indicate that cosmic-ray-derived background, anomaly, and inter-station coherence features provide independent information content beyond standard solar-wind predictors.

The main contributions of this work can be summarized in three aspects. First, we provide empirical evidence that pre-storm cosmic-ray modulation (including FD-like depressions and enhanced inter-station coherence) is systematically associated with subsequent storm-time Dst evolution, supporting its role as a quantitative precursor signal. Second, we establish a practical and reproducible methodology for embedding physically interpretable precursor information into deep learning architectures, thereby bridging data-driven pattern recognition with mechanistic understanding of heliospheric--magnetospheric coupling. Third, the framework is computationally efficient and operationally feasible: under a standard desktop environment, the model produces a single forecast in $\sim$0.8~s, which satisfies real-time requirements for continuous space-weather monitoring and warning.

Several directions can further enhance the proposed framework. Future efforts should explore hybrid architectures that combine sequence models with complementary machine-learning or physics-based components, multi-task learning to jointly predict multiple geomagnetic indices and related space-weather effects, and uncertainty quantification via ensemble or probabilistic forecasting. Expanding the neutron monitor inputs to include additional stations spanning a wider range of geomagnetic cutoff rigidities is also expected to improve disturbance discrimination and robustness for extreme events.

Overall, this work demonstrates that physics-informed machine learning offers a powerful paradigm for space-weather forecasting. By fusing cosmic-ray modulation mechanisms with deep learning, we move toward more accurate, interpretable, and reliable GST prediction systems, which will become increasingly critical as society's dependence on space-based technologies and infrastructure continues to grow.

\begin{acknowledgments}
The authors extend their sincere gratitude to the organizations that provided the essential datasets utilized in this study, including NASA's OMNI database, the Oulu Cosmic Ray Station, the Jungfraujoch Cosmic Ray Station, and the Kyoto University World Data Center for Geomagnetism. Special appreciation is directed toward the research teams and institutions supporting space weather monitoring initiatives worldwide. The work is supported by the National Key R\&D Program of China (2022YFF0503800), the Strategic Priority Research Program of the Chinese Academy of Sciences (XDB0560000), the State Key Program of National Natural Science Foundation of China(12533010), the National Natural Science Foundation of China (12403067, 12350004, 12373111, 12273061, 12273060, 12473095, 12573056, 12503064, 12503062), and the Specialized Research Fund for State Key Laboratory of Solar Activity and Space Weather.
\end{acknowledgments}

\begin{contribution}
Z.G. conceived the idea, designed the research, performed data acquisition, preprocessing, and cosmic-ray feature engineering, developed and trained all LSTM models, conducted the experiments and analysis, produced all figures and tables, and wrote the initial and final drafts of the manuscript.
C.Z. contributed to data visualization, assisted with code optimization and debugging, and participated in result validation.
W.Z. provided valuable suggestions on methodology and helped revise the manuscript.
H.Z. contributed to model optimization and helped revise the manuscript.
G.Z. supervised the entire project, provided critical physical insights into cosmic-ray modulation and GSTs mechanisms, and substantially revised and finalized the manuscript for submission.
All authors discussed the results and commented on the manuscript.
\end{contribution}

\bibliographystyle{aasjournalv7}
\bibliography{refs}

\appendix

\section{Extended Prediction Results}

\subsection{2-Hour Prediction Results}

For 2-hour predictions, RMSE = 5.106~nT.

\begin{figure}[ht!]
\plotone{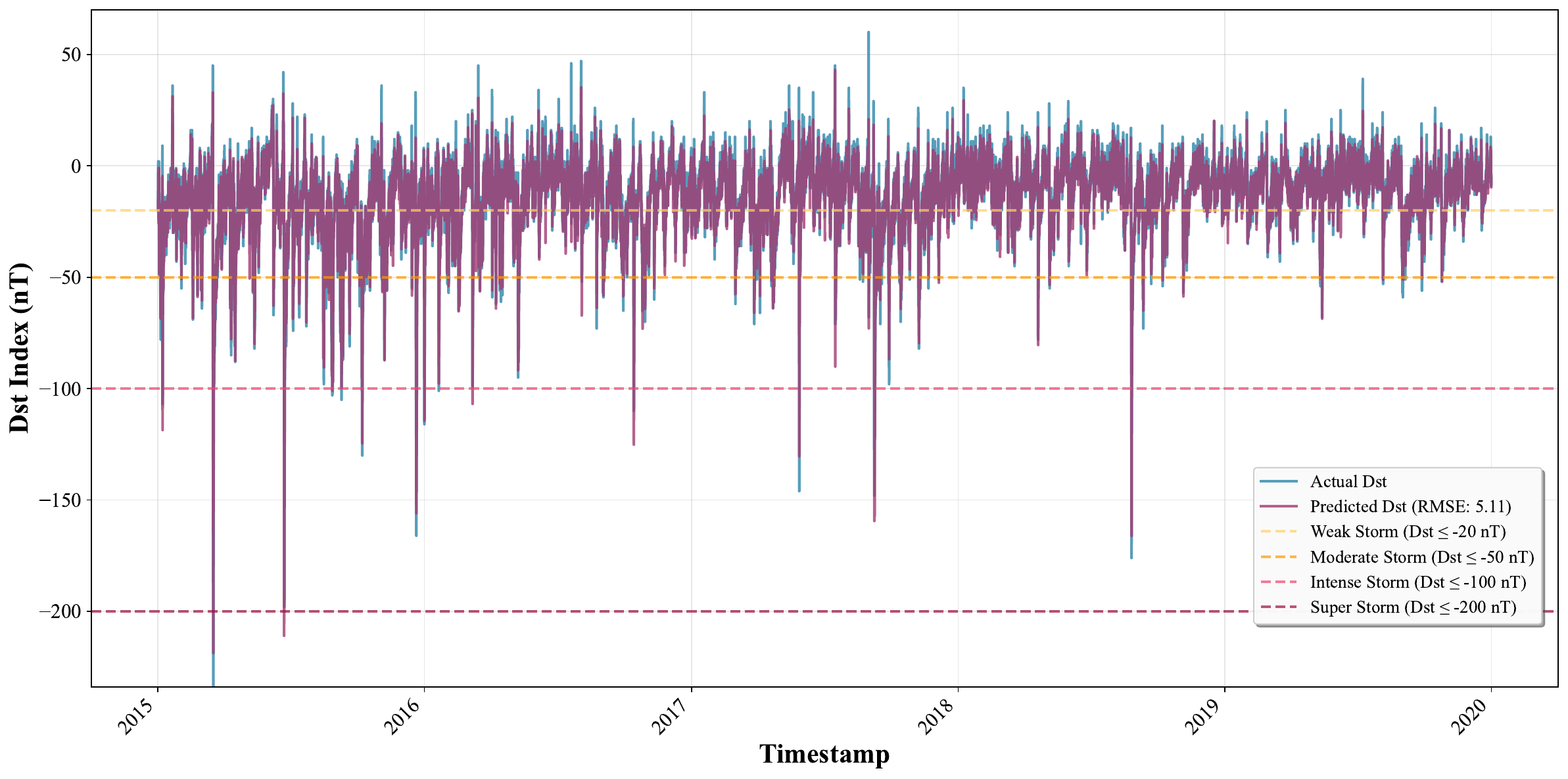}
\caption{2-hour Dst prediction results}
\label{fig:2h}
\end{figure}

\begin{deluxetable}{lccccc}
\tablewidth{0pt}
\tablecaption{2-Hour Statistics and Predictions for Time Points Below Thresholds\label{tab:2h_thresh}}
\tablehead{
\colhead{Classification} & \colhead{Sample Count} & \colhead{Correct Predictions} &
\colhead{Precision} & \colhead{Recall} & \colhead{F1-score}
}
\startdata
Quiet & 32,912 & 31,409 & 0.955 & 0.954 & 0.954 \\
Weak & 9,621 & 7,971 & 0.815 & 0.829 & 0.822 \\
Moderate & 1,162 & 835 & 0.820 & 0.719 & 0.766 \\
Intense & 121 & 94 & 0.832 & 0.777 & 0.803 \\
Super & 3 & 1 & 0.333 & 0.333 & 0.333 \\
\enddata
\tablecomments{Performance metrics for 2-hour Dst index predictions across different geomagnetic activity classifications.}
\end{deluxetable}

\begin{deluxetable}{lccccc}
\tablewidth{0pt}
\tablecaption{2-Hour Statistics and Predictions for GSTs Events\label{tab:2h_event}}
\tablehead{
\colhead{Classification} & \colhead{Sample Count} & \colhead{Correct Predictions} &
\colhead{Precision} & \colhead{Recall} & \colhead{F1-score}
}
\startdata
Weak & 253 & 197 & 0.814 & 0.779 & 0.796 \\
Moderate & 80 & 63 & 0.955 & 0.787 & 0.863 \\
Intense & 12 & 9 & 0.900 & 0.750 & 0.818 \\
Super & 1 & 1 & 0.500 & 1.000 & 0.667 \\
\hline
Overall & 346 & 270 & 0.844 & 0.780 & 0.811 \\
\enddata
\tablecomments{Storm event detection performance for 2-hour forecast horizon.}
\end{deluxetable}

\subsection{12-Hour Prediction Results}

For 12-hour predictions, RMSE = 10.854~nT.

\begin{figure}[ht!]
\plotone{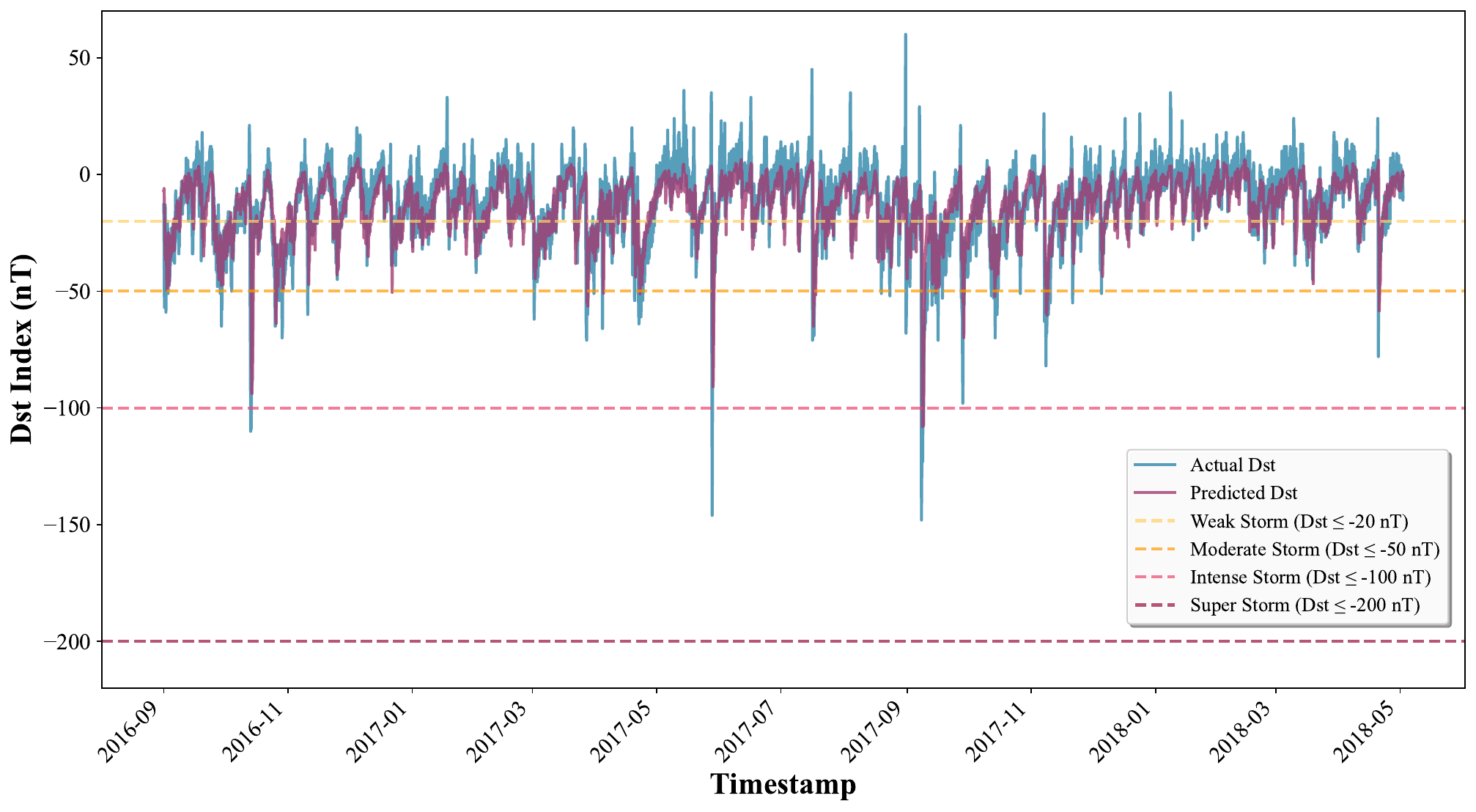}
\caption{12-hour Dst prediction results}
\label{fig:12h}
\end{figure}

\begin{deluxetable}{lccccc}
\tablewidth{0pt}
\tablecaption{12-Hour Statistics and Predictions for Time Points Below Thresholds\label{tab:12h_thresh}}
\tablehead{
\colhead{Classification} & \colhead{Sample Count} & \colhead{Correct Predictions} &
\colhead{Precision} & \colhead{Recall} & \colhead{F1-score}
}
\startdata
Quiet & 32,904 & 31,223 & 0.884 & 0.949 & 0.915 \\
Weak & 9,620 & 5,631 & 0.706 & 0.585 & 0.640 \\
Moderate & 1,162 & 291 & 0.574 & 0.250 & 0.349 \\
Intense & 121 & 8 & 0.500 & 0.066 & 0.117 \\
Super & 3 & 0 & 0.00 & 0.00 & 0.00 \\
\enddata
\tablecomments{Performance metrics for 12-hour Dst index predictions across different geomagnetic activity classifications.}
\end{deluxetable}

\begin{deluxetable}{lccccc}
\tablewidth{0pt}
\tablecaption{12-Hour Statistics and Predictions for GSTs Events\label{tab:12h_event}}
\tablehead{
\colhead{Classification} & \colhead{Sample Count} & \colhead{Correct Predictions} &
\colhead{Precision} & \colhead{Recall} & \colhead{F1-score}
}
\startdata
Weak & 253 & 71 & 0.425 & 0.281 & 0.338 \\
Moderate & 80 & 22 & 0.595 & 0.275 & 0.376 \\
Intense & 12 & 2 & 0.500 & 0.167 & 0.250 \\
Super & 1 & 0 & 0.00 & 0.00 & 0.00 \\
\hline
Overall & 346 & 95 & 0.457 & 0.275 & 0.343 \\
\enddata
\tablecomments{Storm event detection performance for 12-hour forecast horizon.}
\end{deluxetable}

\subsection{24-Hour Prediction Results}

For 24-hour predictions, RMSE = 12.883~nT.

\begin{figure}[ht!]
\plotone{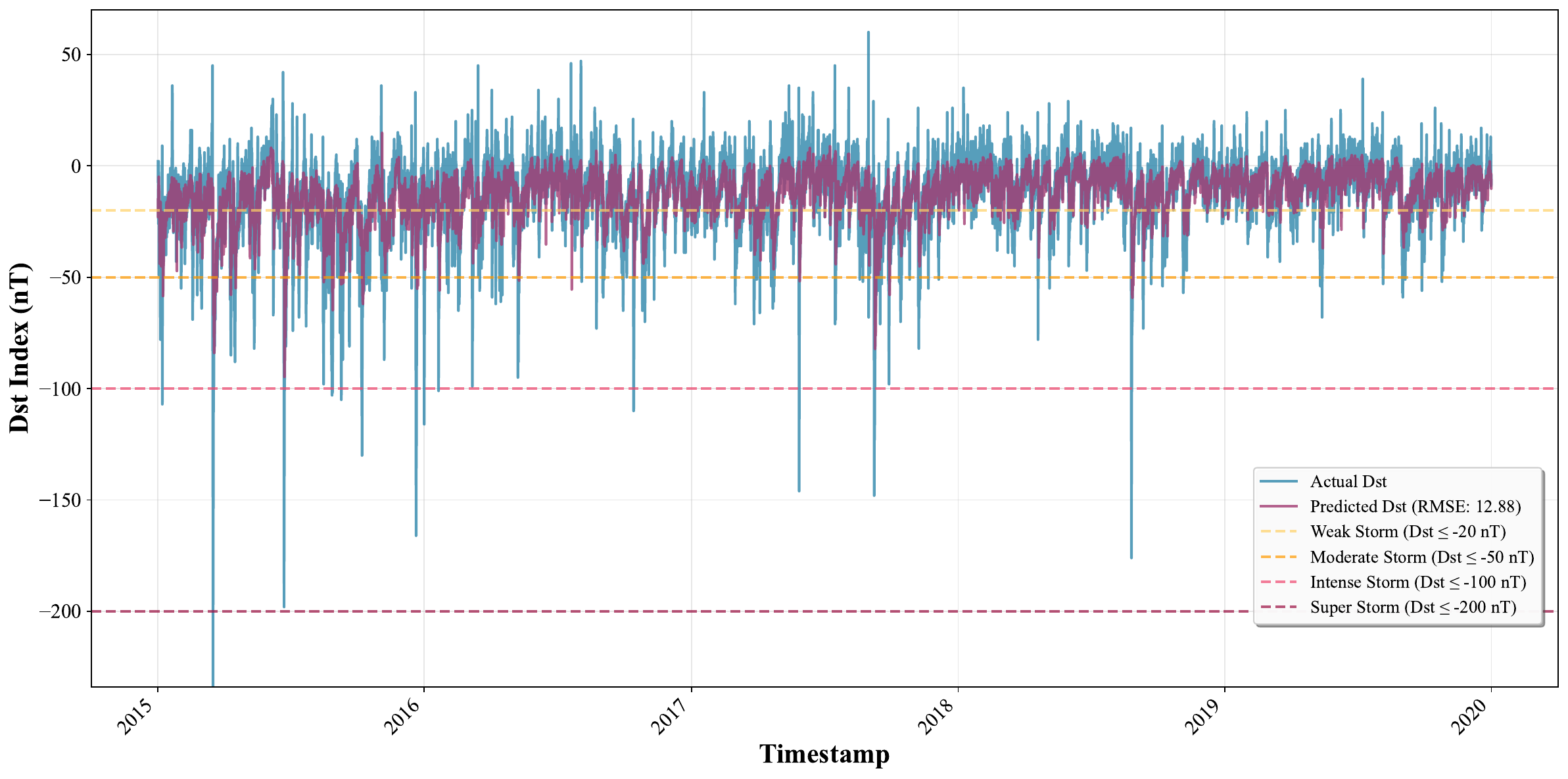}
\caption{24-hour Dst prediction results}
\label{fig:24h}
\end{figure}

\begin{deluxetable}{lccccc}
\tablewidth{0pt}
\tablecaption{24-Hour Statistics and Predictions for Time Points Below Thresholds\label{tab:24h_thresh}}
\tablehead{
\colhead{Classification} & \colhead{Sample Count} & \colhead{Correct Predictions} &
\colhead{Precision} & \colhead{Recall} & \colhead{F1-score}
}
\startdata
Quiet & 32,904 & 30,678 & 0.863 & 0.932 & 0.896 \\
Weak & 9,620 & 5,077 & 0.640 & 0.528 & 0.579 \\
Moderate & 1,162 & 197 & 0.621 & 0.170 & 0.266 \\
Intense & 121 & 0 & 0.00 & 0.00 & 0.00 \\
Super & 3 & 0 & 0.00 & 0.00 & 0.00 \\
\enddata
\tablecomments{Performance metrics for 24-hour Dst index predictions across different geomagnetic activity classifications.}
\end{deluxetable}

\begin{deluxetable}{lccccc}
\tablewidth{0pt}
\tablecaption{24-Hour Statistics and Predictions for GSTs Events\label{tab:24h_event}}
\tablehead{
\colhead{Classification} & \colhead{Sample Count} & \colhead{Correct Predictions} &
\colhead{Precision} & \colhead{Recall} & \colhead{F1-score}
}
\startdata
Weak & 253 & 67 & 0.340 & 0.265 & 0.298 \\
Moderate & 80 & 9 & 0.529 & 0.113 & 0.186 \\
Intense & 12 & 0 & 0.00 & 0.00 & 0.00 \\
Super & 1 & 0 & 0.00 & 0.00 & 0.00 \\
\hline
Overall & 346 & 76 & 0.355 & 0.220 & 0.271 \\
\enddata
\tablecomments{Storm event detection performance for 24-hour forecast horizon.}
\end{deluxetable}

\subsection{48-Hour Prediction Results}

For 48-hour predictions, RMSE = 14.788~nT.

\begin{figure}[ht!]
\plotone{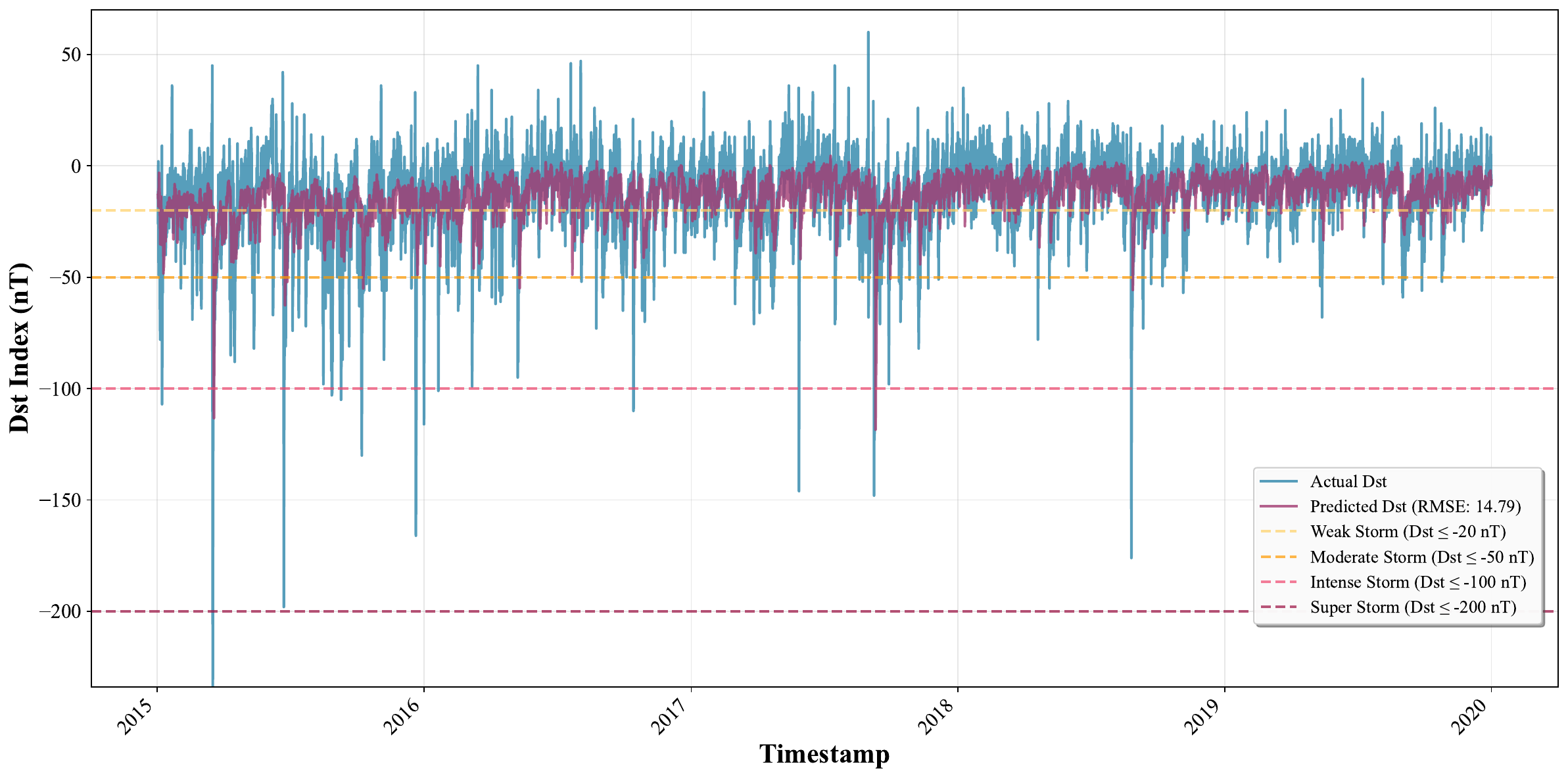}
\caption{48-hour Dst prediction results}
\label{fig:48h}
\end{figure}

\begin{deluxetable}{lccccc}
\tablewidth{0pt}
\tablecaption{48-Hour Statistics and Predictions for Time Points Below Thresholds\label{tab:48h_thresh}}
\tablehead{
\colhead{Classification} & \colhead{Sample Count} & \colhead{Correct Predictions} &
\colhead{Precision} & \colhead{Recall} & \colhead{F1-score}
}
\startdata
Quiet & 32,895 & 30,609 & 0.827 & 0.931 & 0.876 \\
Weak & 9,620 & 3,895 & 0.584 & 0.405 & 0.478 \\
Moderate & 1,162 & 51 & 0.464 & 0.044 & 0.080 \\
Intense & 121 & 0 & 0.00 & 0.00 & 0.00 \\
Super & 3 & 0 & 0.00 & 0.00 & 0.00 \\
\enddata
\tablecomments{Performance metrics for 48-hour Dst index predictions across different geomagnetic activity classifications.}
\end{deluxetable}

\begin{deluxetable}{lccccc}
\tablewidth{0pt}
\tablecaption{48-Hour Statistics and Predictions for GSTs Events\label{tab:48h_event}}
\tablehead{
\colhead{Classification} & \colhead{Sample Count} & \colhead{Correct Predictions} &
\colhead{Precision} & \colhead{Recall} & \colhead{F1-score}
}
\startdata
Weak & 253 & 54 & 0.325 & 0.213 & 0.258 \\
Moderate & 80 & 3 & 0.600 & 0.037 & 0.071 \\
Intense & 12 & 1 & 0.500 & 0.083 & 0.143 \\
Super & 1 & 0 & 0.00 & 0.00 & 0.00 \\
\hline
Overall & 346 & 58 & 0.335 & 0.168 & 0.224 \\
\enddata
\tablecomments{Storm event detection performance for 48-hour forecast horizon. The integration of cosmic-ray features improves the F1-score by 25.84\% compared to models without these features.}
\end{deluxetable}

\section{Detailed LSTM Model Analysis}

\subsection{LSTM Architecture}

This appendix provides an in-depth analysis of the LSTM neural network utilized for forecasting the Dst index, supplementing the concise overview presented in Section 4.3. The LSTM architecture is selected for its capability to model long-term dependencies in time-series data, a critical attribute for capturing the multi-phase evolution of GSTs (initial, main, and recovery phases).

\begin{figure}[ht!]
\plotone{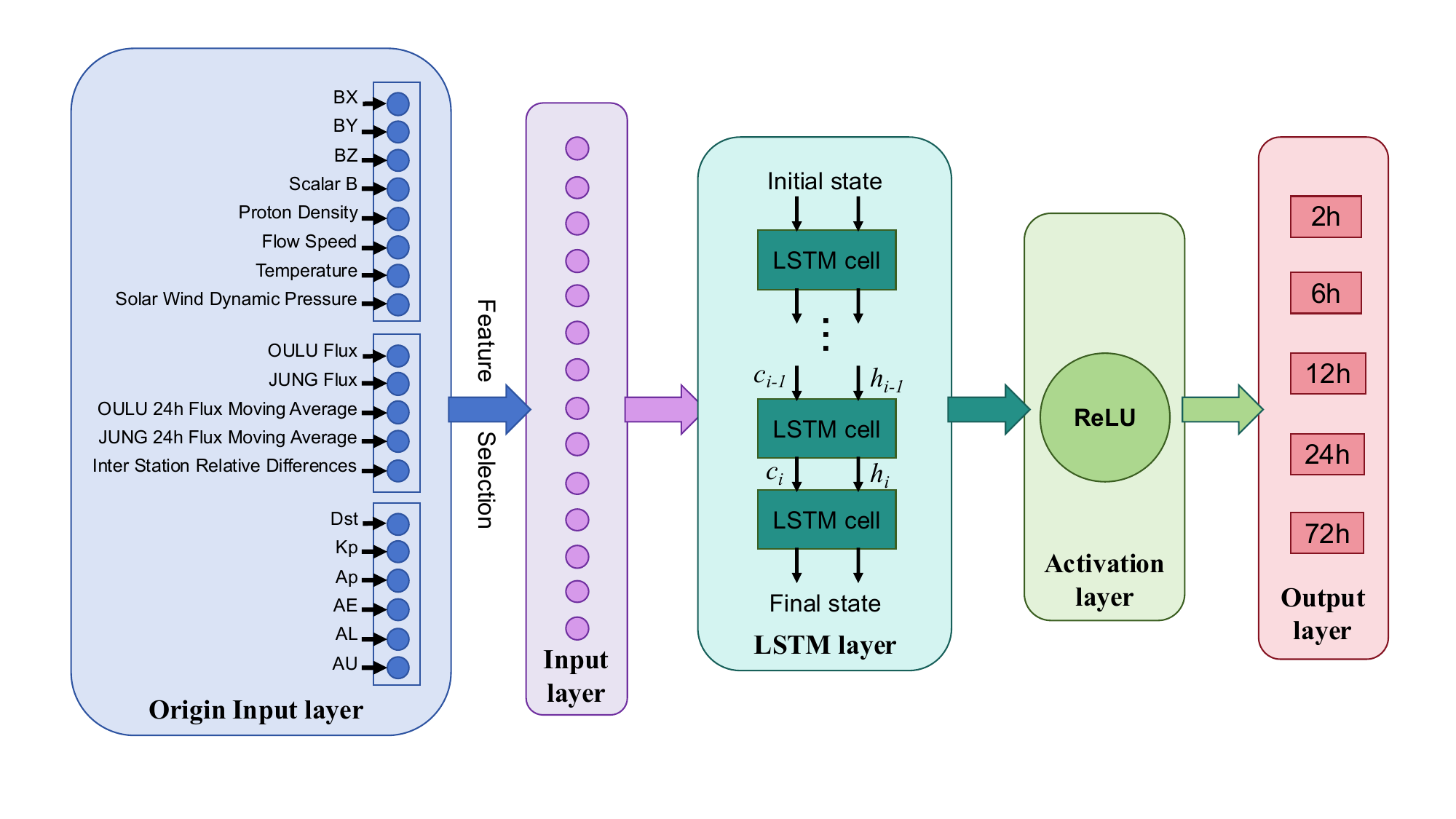}
\caption{Flowchart of LSTM Architecture for Dst Index Prediction}
\label{fig:lstm_arch}
\end{figure}

The LSTM model is implemented as a sequential neural network comprising a single LSTM layer with 50 units, followed by a dense output layer with one neuron to produce the Dst forecast. The LSTM layer employs the Rectified Linear Unit (ReLU) activation function to introduce non-linearity. The input shape is defined as (sequence length, 19), where the sequence length corresponds to the specific forecast target.

Feature selection is informed by permutation-based importance analysis, which identifies critical inputs including the BZGSM component, solar wind dynamic pressure, and cosmic-ray anomalies. Correlation analysis further refined the feature set, reducing it from 19 to 16 by excluding highly correlated features ($|r| > 0.85$) and low-importance features (importance $< 0.002$), such as the AL index and solar wind plasma temperature, to improve computational efficiency without sacrificing predictive accuracy.

\begin{figure}[ht!]
\plotone{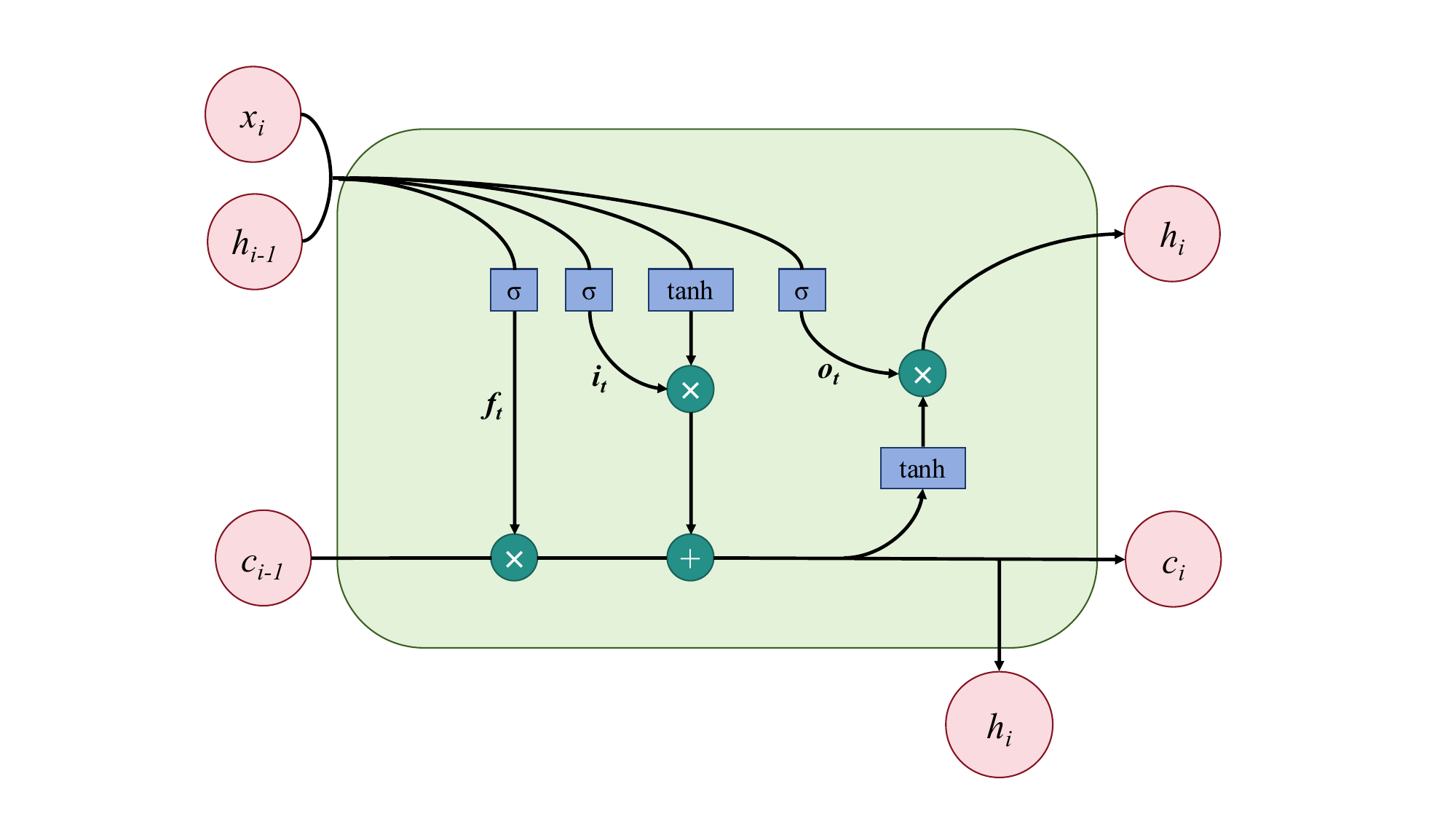}
\caption{LSTM cell structure and gating mechanisms}
\label{fig:lstm_cell}
\end{figure}

The model's predictive performance is quantified by RMSE values of 5.106~nT, 8.315~nT, 10.854~nT, 12.883~nT, and 14.788~nT for the 2-hour, 6-hour, 12-hour, 24-hour, and 48-hour forecast horizons. The integration of cosmic-ray features significantly improves storm event detection, evidenced by a 25.84\% enhancement in the 48-hour F1-score compared to models without these features. These results underscore the efficacy of combining physics-informed inputs with deep learning techniques.

\end{document}